\documentclass[sigconf]{acmart}

\AtBeginDocument{%
  \providecommand\BibTeX{{%
    \normalfont B\kern-0.5em{\scshape i\kern-0.25em b}\kern-0.8em\TeX}}}

\copyrightyear{2025}
\acmYear{2025}
\setcopyright{acmcopyright}
\acmConference[Conference acronym 'XX]{Make sure to enter the correct
  conference title from your rights confirmation email}{June 03--05,
  2018}{Woodstock, NY}
\acmPrice{15.00}
\acmDOI{10.1145/3491102.3501940}
\acmISBN{978-1-4503-9157-3/22/04}

\usepackage{_macros}
\usepackage{caption}
\usepackage{subcaption}
\usepackage{array}
\usepackage{tikz}
\usepackage{hyperref}
\usepackage{hyperxmp}

\newcommand*\circled[1]{\tikz[baseline=(char.base)]{
            \node[shape=circle,draw,inner sep=1pt] (char) {#1};}}

\newcommand{\name}{TherAIssist}

\begin{document}

\title[\name: Assisting Art Therapy through Human-AI Interaction]{\name: Assisting Art Therapy Homework and Client-Practitioner Collaboration through Human-AI Interaction}

\author{Di Liu}
\authornote{Both authors contributed equally to this research.}
\orcid{0009-0001-3513-6587}
\affiliation{%
  \institution{School of Design, SUSTech}
  \city{Shenzhen}
  \country{China}
}
\email{seucliudi@gmail.com}

\author{Jingwen Bai}
\authornotemark[1]
\orcid{0000-0001-5118-993X}
\affiliation{%
  \institution{School of Computing, National University of Singapore}
  \streetaddress{13 Computing Drive}
  \city{Singapore}
  \country{Singapore}
  \postcode{117417}
}
\email{levanukp430@gmail.com}

\author{Zhuoyi Zhang}
\orcid{0009-0001-0308-6861}
\affiliation{%
  \institution{Department of Human Centered Design \& Engineering, University of Washington}
  \city{Seattle}
  \country{UnitedStates}
}
\email{zhuoyiz@uw.edu}

\author{Yilin Zhang}
\orcid{0009-0002-0675-1545}
\affiliation{%
  \institution{Carnegie Mellon University}
  \city{Pittsburgh}
  \country{United States}
}
\email{yilinjz@cmu.edu}

\author{Zhenhao Zhang}
\orcid{0009-0001-9242-9618}
\affiliation{%
  \institution{School of Design, SUSTech}
  \city{Shenzhen}
  \country{China}
}
\email{zhenhao_z@163.com}

\author{Jian Zhao}
\orcid{0000-0001-5008-4319}
\affiliation{%
  \institution{University of Waterloo}
  \city{Waterloo}
  \country{Canada}
}
\email{jianzhao@uwaterloo.ca}

\author{Pengcheng An}
\authornote{Corresponding Author.}
\orcid{0000-0002-7705-2031}
\affiliation{%
  \institution{School of Design, SUSTech}
  \city{Shenzhen}
  \country{China}
}
\email{anpc@sustech.edu.cn}



\begin{abstract}

Art therapy homework is essential for fostering clients' reflection on daily experiences between sessions. However, current practices present challenges: clients often lack guidance for completing tasks that combine art-making and verbal expression, while therapists find it difficult to track and tailor homework. How HCI systems might support art therapy homework remains underexplored. To address this, we present TherAIssist, comprising a client-facing application leveraging human-AI co-creative art-making and conversational agents to facilitate homework, and a therapist-facing application enabling customization of homework agents and AI-compiled homework history. A 30-day field study with 24 clients and 5 therapists showed how TherAIssist supported clients’ homework and reflection in their everyday settings. Results also revealed how therapists infused their practice principles and personal touch into the agents to offer tailored homework, and how AI-compiled homework history became a meaningful resource for in-session interactions. Implications for designing human-AI systems to facilitate asynchronous client-practitioner collaboration are discussed.

\end{abstract}



\begin{teaserfigure}
  \includegraphics[width=\textwidth]{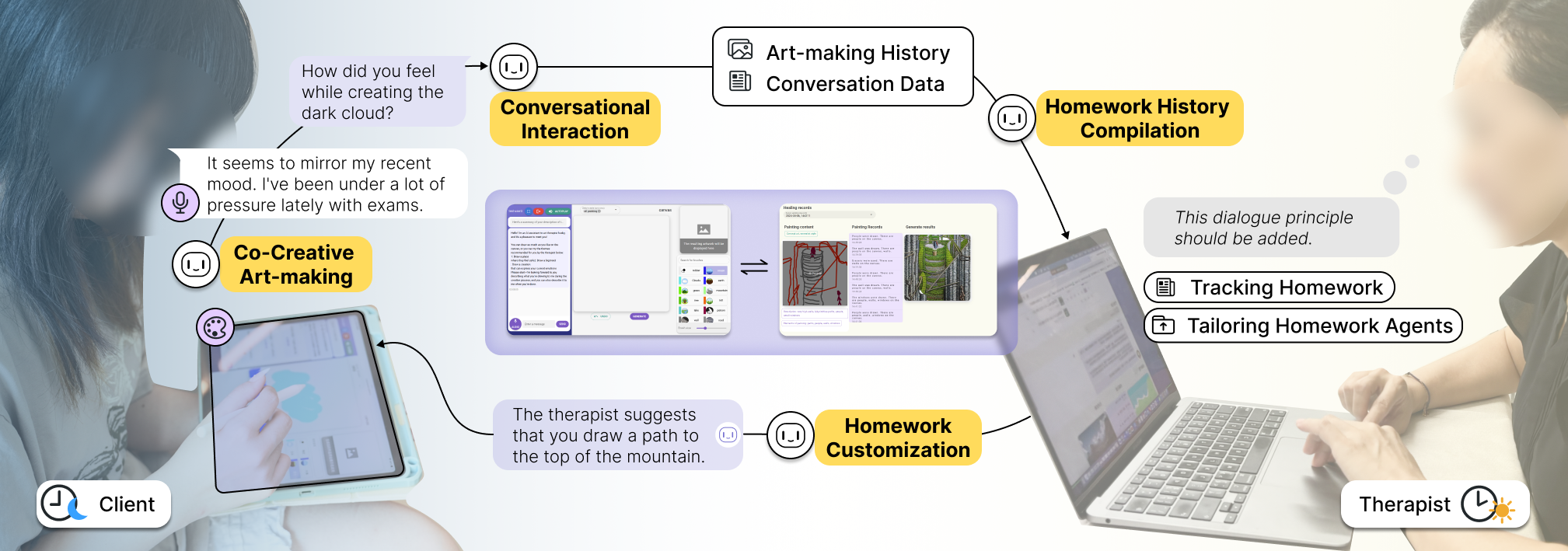}
  \vspace{-7mm}
  \caption{\name{} enables clients to leverage human-AI co-creative art-making and conversational agents to facilitate art therapy homework; Meanwhile, it can support therapists customizing homework agents and reviewing AI-compiled homework history.}
  \Description{ Figure 1 illustrates the interaction flow of our system. On the left, the person represents our clients, while on the right, the person represents our therapists. The client on the left is using a tablet to complete their homework through our system, drawing while interacting with a conversational agent. The art-making history and conversation data are simultaneously transmitted to the therapist-facing application. The therapist can track this homework history and customize homework agents for the client. These homework assignments are then sent through the system to the client-facing application. The overall theme of the image is emphasized through the use of yellow and black, highlighting this asynchronous collaboration.
}
  \label{fig:teaser}
\end{teaserfigure}


\maketitle

\section{Introduction}
Art therapy is a multi-modal therapeutic approach that combines creative activities with verbal expression to help individuals articulate their feelings and enhance their mental well-being~\cite{malchiodi2011handbook}.
It includes both \textit{in-session} and \textit{between-session} activities.
In-session therapy involves conversation and collaborative work between the therapist and the client.
Meanwhile, between-session activities---referred to as ``\textit{therapy homework}'', play a vital role by incorporating assignments that enable clients to carry the therapeutic work into their everyday lives~\cite{kazantzis2007handbook}. 
It has been proposed that ``homework'' is a common element in psychotherapy, including art therapy~\cite{heckwolf2014coordinating}, cognitive-behavioral therapies~\cite{kazantzis2007handbook}, solution-focused therapy~\cite{beyebach1996research}, self-help interventions~\cite{jordan1995programmed}, and others.
Therapy homework in psychotherapy in general, serves the common goals of helping clients reinforce therapy progress through practice in natural environments~\cite{beerse2020therapeutic}, while also strengthening their positive relationship with the therapists~\cite{huckvale2009case}.
These therapy homework in psychotherapy typically include art-making~\cite{hoshino2011narrative,barrow2020experiences}, journaling~\cite{borkin2014healing,smith2019visual}, emotion-self report~\cite{nesset2021does} as well as mindfulness and relaxation techniques~\cite{knapp2024daily}.

In art therapy, it is common for art therapists to assign homework to clients~(see \autoref{fig:context1})~\cite{hoshino2011narrative}.
They often recommend art therapy homework that integrates art-making with verbal expression~\cite{smith2019visual,heckwolf2014coordinating,hoshino2011narrative}.
This combination allows clients to explore their innermost thoughts and find a cathartic outlet through verbal expression, while also benefiting from the creative and expressive nature of art-making~\cite{smith2019visual}. 
However, without a therapist's guidance, the threshold for therapeutic art-making can increase~\cite{du2024deepthink}, and clients in psychotherapy often struggle with constructing a narrative, understanding their past, and verbalizing their feelings and experiences~\cite{pennebaker1999forming,mayer2011emotional}.

In the field of human-computer interaction (HCI), limited knowledge has been accumulated about how to effectively support between-session therapy activities~\cite{Oewel_2024}, including art therapy homework that uniquely integrates art-making with verbal expression. 
A few recent cases suggest that human-AI co-creative approach can lower the barrier to art-making and enhance creativity in therapy practices~\cite{du2024deepthink, liu2024he}. 
Extensive research also highlights how conversational agents can aid clients in self-exploration, self-disclosure, and emotional support~\cite{kim2024mindfuldiary, park2021wrote, jo2023understanding, seo2024chacha}. 
Yet, so far, little empirical research exists on combining human-AI co-creative art-making with conversational agents to support clients' art therapy homework.


Therapists also face challenges in tracking and tailoring art therapy homework. Homework fosters an asynchronous therapist-client collaboration that keeps clients engaged between sessions~\cite{huckvale2009case,hoshino2011narrative} and builds long-term trust and intimacy~\cite{huckvale2009case}.
This asynchronous collaboration requires therapists to track homework records and history, which serve as valuable data for understanding clients' thoughts between sessions and can help shape the treatment approach~\cite{kazantzis2007handbook,hoshino2011narrative,richards2018impact}.
However, managing and tracking homework records can be difficult and often increases therapists' workload~\cite{richards2018impact}.
Moreover, prior studies also highlight the importance of therapists tailoring therapy homework while tracking its progress, which can reinforce that therapeutic process~\cite{Oewel_2024, kazantzis2007handbook}. 
However, assisting therapists in creating structured homework that is tailored to the client's needs and incorporates emotional support remains a significant challenge~\cite{kazantzis2007handbook, coon2002encouraging, Oewel_2024}.

Outside of art therapy, several HCI studies have explored how AI conversational agents can assist therapists by summarizing and presenting relevant health information in asynchronous collaborations between clients and practitioners~\cite{kim2024mindfuldiary, yang2024talk2care,10.1145/3659604}. 
For instance, MindfulDiary uses an LLM-based dashboard to help clinicians empathize with patients and understand their daily thoughts~\cite{kim2024mindfuldiary}. 
Additionally, beyond the health domain, recent research has investigated no-code design tools that allow users to customize conversational agents with tailored dialogue flows or styles~\cite{hedderich2024piece, ha2024clochat}. 
Nonetheless, how to support therapist-client asynchronous collaboration surrounding art therapy homework remains an unaddressed opportunity.



Building upon these motivations, we present \name{} (\autoref{fig:teaser}), a human-AI system consisting of: 
(1) a client-facing application that combines human-AI co-creative art-making with conversational interaction to facilitate clients' art therapy homework in their daily settings; 
and (2) a therapist-facing application that features AI-compiled homework history and customization of client homework agents to offer tailored homework guidance.


Taking \name{} as both a novel system to study about, and a research tool to study with, we conducted a field deployment involving 24 recruited clients and five therapists, over a period of one month. We set out to explore the following research questions:

\begin{itemize}
  \item \textbf{\textit{RQ1: how would a human-AI system support clients' art therapy homework in their daily settings?}}
  \item \textbf{\textit{RQ2: how would a human-AI system mediate therapist-client collaboration surrounding art therapy homework?}}
\end{itemize}

The field evaluation yielded rich empirical data, revealing how \name{} facilitated clients' homework practice across various natural environments. It also offered diverse examples of how combining conversational interaction with art-making supported clients in articulating their feelings and experiences, potentially enabling them to explore the deeper meaning behind their artwork. 
Moreover, the data illustrate how therapists customized homework agents to embed their professional beliefs and personal style, making the agents reflect their unique approach and personal touch. The findings also depict how therapists leveraged the AI-compiled homework history, including both artworks and dialogues, to better understand clients' characteristics and emotional state, identify triggers for deeper discussion, and empower clients during in-session therapy.


This work thereby contributes twofold: (1) a novel system for art therapy homework; and (2) rich empirical findings contextualizing how a human-AI system facilitated clients' homework in daily settings and client-practitioner collaboration, with relevant HCI design implications discussed.

\section{Background and Related Work} \label{sec:background}
\subsection{Art Therapy and Therapy Homework}
Art therapy, guided by a professional art therapist, is a multimodal therapeutic approach that integrates the nonverbal language of art with verbal communication to promote personal growth, insight, and transformation~\cite{malchiodi2007art}.
It includes both in-session and between-session activities. 
In-session activities involve direct interaction between the therapist and client, while between-session activities, often referred to as ``therapy homework'', are personalized exercises assigned by therapists for clients to complete in their real-world environments~\cite{kazantzis2007handbook}.
It has been proposed that ``homework'' may be considered a common element in psychotherapy ~\cite{kazantzis2007handbook}. 
The use of homework is widely applied in many psychotherapy contexts, including art therapy~\cite{kazantzis2000homework, huckvale2009case}, solution-focused therapy~\cite{beyebach1996research}, personal construct therapy~\cite{kazantzis2007handbook}, and so on.
The therapy homework comes in various forms, including journaling~\cite{hoshino2011narrative,dattilio2012collaboration}, drawing~\cite{hoshino2011narrative}, photographing~\cite{huckvale2009case}, workbooks and worksheets~\cite{riley2003family,Oewel_2024}, collages~\cite{riley2003family}, recording~\cite{Oewel_2024}, mindful practices~\cite{smith2019visual}, and photo-elicitation reflective writing~\cite{davis2015mindful}.
Thus, it can help clients practice reflecting on and validating their own feelings and experiences~\cite{riley2003family}. 
Most importantly, it can bring therapy into the ``real'' world, fostering the development of therapeutic skills across various contexts~\cite{kazantzis2007handbook}.
On the other hand, therapy homework helps strengthen the therapeutic alliance between the therapist and client over time, fostering trust and deepening their relationship~\cite{huckvale2009case,cronin2015integrating,sezaki2000home}.

In art therapy, it is \textit{``not unusual''} for therapists to assign homework to clients~\cite{hoshino2011narrative}.
Art therapists commonly assign a variety of therapy homework exercises to clients, often consisting of visual art creating and verbal reflections~\cite{huckvale2009case,smith2019visual,davis2015mindful,hoshino2011narrative}. 
For example, Huckvale et al. reported that therapists encouraged clients to complete therapeutic work such as looking at, drawing, photographing, and writing about the sky every day~\cite{huckvale2009case}.
Smith et al. also demonstrated that therapists could combine art-making with verbal exercises in homework assignments to enhance clients' daily coping mechanisms, which has been proven to be effective for both patients and caregivers~\cite{smith2019visual}.
This combination of art-making and verbal expression is not only applied in art therapy but also incorporated by other therapeutic contexts~\cite{peretz2023machine,kazantzis2007handbook}. It can leverage the benefits of exploring innermost thoughts and new meaning through verbal expression while harnessing the creative and expressive qualities inherent in the art-making process~\cite{smith2019visual}. 

Despite its benefits, art therapy homework faces several challenges.
Namely, without a therapist's guidance, engaging in therapeutic art-making can be even more challenging, leading to frustration ~\cite{du2024deepthink}.
Further, clients in psychotherapy often find it difficult to build a narrative and express their emotions and experiences verbally~\cite{pennebaker1999forming,mayer2011emotional,kim2024mindfuldiary}. Kazantzis et al. found that practicing without structure or support can trigger clients' negative beliefs and create practical barriers~\cite{kazantzis2022comprehensive}.

Meanwhile, assisting therapists in tailoring homework remains a significant challenge.
Therapists need to tailor therapy homework to align with the clients' current state and capacity~\cite{Oewel_2024}. 
Good tailoring can help clients boosts their motivation to engage with homework and provides the connections from one session to the next~\cite{Oewel_2024,katz2023assigning}.
However, designing well-structured homework tasks tailored to clients' needs can be challenging for therapists, potentially leading to difficulties in gathering essential client data, or creating discomfort~\cite{kazantzis2007handbook, Oewel_2024}.
It is also challenging for therapists to provide consistent encouragement between sessions, which can lead to low homework compliance~\cite{dryden2011cbt}.

Furthermore, effectively tracking clients' homework is also difficult.
The homework provides an opportunity for therapists to collect valuable data, which they can explore, analyze, and synthesize to gain deeper insights into the client's progress~\cite{kazantzis2007handbook}.
The homework data can help therapists better understand clients' issues, uncover their strengths, and shape a more effective treatment approach~\cite{kazantzis2007handbook,gereb2022online}.
However, therapists could face difficulties in adequately tracking homework history, as it increases their workload and homework data cannot be easily organized and revisited~\cite{richards2018impact, gereb2022online,peretz2023machine}.
Ongoing therapeutic collaboration is a key to enhance homework adherence and foster positive relationships between therapists and clients~\cite{thomas2008evaluating, Ali2017_face2emoji}.

\subsection{Digital Art Therapy and Human-AI Co-creation}
Research has explored how digital technologies could enhance accessibility, engagement, and collaboration for in-session art therapy. Various digital tools, such as virtual reality~\cite{kaimal2020virtual}, online chat applications~\cite{collie2006distance, hankinson2022keeping}, digital art-making tools~\cite{darewych2015digital, choe2014exploration}, and specialized art therapy systems~\cite{cubranic1998computer,yilma2024artful}, have been examined for their therapeutic potential. 
For instance, Collie et al. developed a computer-supported distanced art therapy system that enables audio and visual communication in individual or group sessions~\cite{collie2002computer}. 
Digital art therapy has shown promise in bridging geographical gaps~\cite{levy2018telehealth, collie2017online}, increasing accessibility for clients facing stigma or disabilities~\cite{kim2023case}, boosting engagement in creative processes~\cite{levy2018telehealth}, and fostering therapeutic rapport~\cite{collie2002computer, orr2012technology}. 
However, most of these studies have focused on in-session therapy. While the benefits of digital tools for in-session therapy are well-documented by these studies, between-session scenarios remain under-explored. 



Recent advancements in Generative Artificial Intelligence (GenAI) present promising opportunities for human-AI co-creative approaches which could lower the threshold to creative processes in therapeutic settings ~\cite{sun2024understanding, wan2024metamorpheus,jutte2024perspectives, du2024deepthink,liu2024he}. For instance, Sun et al. found that human-AI collaboration in music therapy can improve therapeutic efficiency, reduce the complexity of music creation, and increase client engagement~\cite{sun2024understanding}. Similarly, DeepThInk, an Generative Adversarial Network (GAN) based system, has shown that co-creative art-making with AI can lower the threshold for artistic expression in art therapy~\cite{du2024deepthink}. 
Nonetheless, these studies mainly encompassed therapeutic activities where therapists provided real-time guidance. While DeepThInk also suggested promise for asynchronous use, it only supported art-making without verbal expression. Despite this, human-AI co-creative approach holds potential for art therapy homework. Building on this, our work explores integrating this approach with conversational interaction to support art therapy homework.

\subsection{AI Agents for Mental Health}

Despite limited research on how AI conversational agents support art therapy homework, a substantial body of HCI research has explored the design of rule-based and retrieval-based conversational agents for various therapeutic techniques beyond art therapy. These studies have examined applications in areas such as self-disclosure and self-compassion~\cite{lee2020designing, park2021designing, lee2019caring}, problem-solving therapy~\cite{o2018suddenly, kannampallil2023effects}, expressive writing~\cite{park2021wrote}, mindfulness practices~\cite{seah2022designing, inkster2018empathy}, positive psychology~\cite{kannampallil2023effects, jeong2023deploying}, and cognitive behavioral therapy~\cite{fulmer2018using, fitzpatrick2017delivering, su2020analyzing}. For instance, Diarybot used conversational agents to facilitate expressive writing, successfully encouraging participants to share emotions and stories~\cite{park2021wrote}. These studies highlight initial successes and provide valuable insights for designing AI conversational agents to support structured mental health interventions. However, technological limitations---such as challenges in understanding conversational context and restricted linguistic capabilities---can lead to unnatural or irrelevant interactions, reducing user engagement~\cite{ma2024evaluating,kim2024mindfuldiary}.


The recent advancements in large language models~(LLMs) have resulted in remarkable breakthroughs in developing dialogue systems that are more naturalistic and adaptive~\cite{bae2022building,hamalainen2023evaluating}. 
Recent studies in HCI have explored LLMs were leveraged to support various aspects of mental health, such as promoting cognitive reframing of negative thoughts~\cite{sharma2024facilitating, sharma2023cognitive}, facilitating sharing of emotions and experiences~\cite{seo2024chacha}, enhancing mindfulness~\cite{kumar2023exploring}, providing ad-hoc mental health support for specific groups~\cite{ma2024evaluating}, and enabling context-aware journaling~\cite{nepal2024contextual,10.1145/3699761}.
For instance, ChaCha, an LLM-driven conversational system, tracks conversation context to help children express their stories and emotions while identifying their feelings related to positive and negative events~\cite{seo2024chacha}. These studies offer valuable insights into how LLM-based conversational agents can facilitate deep self-disclosure and emotional support without direct therapist involvement.

Less studies explored how AI agents can support asynchronous client-practitioner collaboration in health-related domains. Some have examined AI's role in summarizing and presenting important health information from users' dialogue history to assist therapists~\cite{yang2024talk2care, kim2024mindfuldiary,10.1145/3659604}. For example, MindfulDiary, a LLM-powered journaling app, helps psychiatric patients document daily experiences and provides a clinician dashboard for practitioners to review entries~\cite{kim2024mindfuldiary}. 
However, these studies have not yet addressed how therapists can dynamically tailor AI agents to better meet their clients' mental health needs.

Yet outside of health domains, there is growing interest in designing no-code tools that allow users to customize conversational agents by creating personalized dialogue flows and styles~\cite{zheng2023synergizing, ha2024clochat, hedderich2024piece, Bhattacharjee2024}. For instance, Michael et al. designed a no-code chatbot design tool that lets users modify conversation flows to assist in bystander education~\cite{hedderich2024piece}. Yet, understanding how to customize AI agents for art therapy and mediate asynchronous therapist-client collaboration surrounding therapy homework remains underexplored, which has motivated our study.

\section{Contextual Understanding}
Recent HCI research has pinpointed the significance of understanding therapy homework in mental health~\cite{Oewel_2024}, yet art therapy homework remains a unique and unaddressed domain. 
Therefore, we conducted a contextual study with a group of therapists to gain a concrete understanding of current art therapy homework practice and to identify common needs for technological support.
\begin{figure*}[tb]
  \centering
  \includegraphics[width=\linewidth]{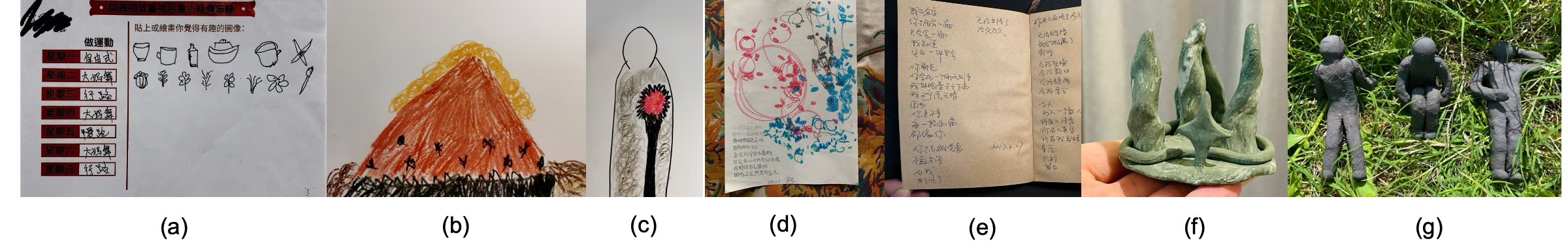}
  \vspace{-4mm}
  \caption{Art therapy homework outcomes from the therapists' previous practice: (a): T4; (b)-(c): T5; (d)-(g): T3}
  \Description{This Figure showcases the outcomes of homework practices among art therapists. From left to right: (a) a client completing a homework task on a structured worksheet; (b) a depiction of a volcano represented by yellow patterns, with orange indicating imminent erupting lava; (c) an outline of a small figure containing a floral pattern in black and red; (d) a composition using text alongside red and blue floral designs; (e) a diary entry documented by a client; (f) a handcrafted green mountain created by a client; (g) a client-made black clay figurine placed on a patch of grass.}
  \label{fig:context1}
\end{figure*}

\begin{table*}[tb]
\caption{Demographics of Participant Therapists: Experience refers to the number of years engaged in art therapy; The Number of Case refers to cases related to art therapy; The Number of Online Case refers to cases related to online art therapy}
\label{tab:expert}
\vspace{-3mm}
\small
\resizebox{\textwidth}{!}{
\begin{tabular}{ccccccccc}
\toprule
ID & Age & Gender & Experience & Education Level& Major & Region & The Number of Case&The Number of Online Case\\
\midrule
T1& 39& F& 6 & Master & Art Therapy & United States(Florida) &300+&65\\
T2& 41& F& 10 & Master & Art Therapy & Italy(Puglia) &200+&12\\
T3& 49& F& 8 & Master & Art Therapy & China(Guangdong) &350+&85\\
T4& 37& F& 5 & PhD & Cognitive Psychology\&Art Therapy&China(Hongkong) &100+&52\\
T5& 24& F& 2 & Master & Art Therapy& China(Hangzhou) &100&45\\
\bottomrule
\end{tabular}
}
\Description{Table 1 presents the demographics of the participant therapists. Experience refers to the number of years they have been engaged in art therapy, and the Number of Cases indicates the number of art therapy cases they have handled. The five therapists are as follows: T1 is 39 years old, female, with 6 years of experience. She holds a Master’s degree in Art Therapy and practices in Florida, United States, having managed over 300 cases. T2 is 41 years old, female, with 10 years of experience. She has a Master’s degree in Counseling Psychology and works in Puglia, Italy, with more than 200 cases. T3 is 49 years old, female, with 8 years of experience. She holds a Master’s degree in Art Therapy and is based in Guangdong, China, having overseen over 350 cases. T4 is 37 years old, female, with 5 years of experience. She has a PhD in Cognitive Psychology and Art Therapy and practices in Hong Kong, China, having handled more than 100 cases. T5 is 24 years old, female, with 2 years of experience. She holds a Master’s degree in Art Therapy and is located in Hangzhou, China, with approximately 100 cases managed. The Number of Online Case refers to cases related to online art therapy
}
\end{table*}

\subsection{Procedure and Preparation}
Five art therapists (T1-T5; all self-identified females; aged 24-49) participated in this study. None of the therapists were members of the research team. T3 was a previous collaborator; the other therapists were recruited via T3's professional network, intended for a diverse group of practitioners from various geographical locations.
Their demographics and expertise are detailed in~\autoref{tab:expert}.
We first conducted 60-minute remote one-on-one interviews with each therapist to understand their current homework practice. This was followed by two 60-minute online focus groups with the therapists. The researchers, acting as facilitators, moderated the discussion on the common challenges for homework practice, aiming to identify needs and design opportunities.
In addition, we kept close collaboration with the therapists throughout the development phase and conducted informal follow-ups to gather inputs in formulating design features.
The one-on-one interviews and focus group sessions were screen-recorded and transcribed. We conducted open coding and affinity diagramming to identify emerging insights reported below.

\subsection{Contextual Understanding: Current Practice and Common Challenges}

Our therapists confirmed that art therapy homework plays a crucial role in helping them understand and collaborate with clients between sessions. They shared their current methods for assigning art therapy homework, which often involves multi-modal activities~(see \autoref{fig:context1}) combining visual arts (e.g., drawing, collage-making, photography and clay sculpting) with written or spoken documentation of emotions and experiences (e.g., journaling, social media posts, and audio recordings).
The therapists noted that integrating visual presentations with verbal expression is a common practice, as it helps clients document and articulate their experiences. For example, T4 combined art-making with audio recording to assist clients in expressing their current feelings: \qt{I asked the elderly [clients] to take photos and create collages at home and encouraged them to record audio to share their daily emotions}. The therapists believed that this combination encourages clients to more fully describe their artwork, explore subconscious thoughts behind the creative process, and gain new perspectives on their lives.
Aside from their approach of leveraging art therapy homework in current practice, the therapists also share their challenges regarding art therapy homework. From their shared experiences, three major sets of challenges emerged, which are summarized below:

\subsubsection{\textbf{CH1}: Challenges in Homework Threshold without Therapist Guidance} 

Our therapists indicated that art-making-based therapy homework can pose a creative barrier for clients without therapist guidance~(\textbf{CH1-1}). T4 noted that this barrier could lead to stress, self-criticism, and fear of failure: \qt{If a client is self-critical, they may fear creating something `ugly', which can increase pressure and hinder the therapeutic process}. Consistent with prior studies~\cite{Tang2017,Harwood2007}, the therapists also confirmed that clients may lack confidence in completing homework or producing emotional responses without guidance, which can result in lower compliance.
Additionally, therapists expressed concerns that clients might struggle to interpret their artwork in a therapeutic way without support, reducing their motivation for deep reflection~(\textbf{CH1-2}). T1 observed that without proper guiding, it can be difficult for clients to make full use of the exercise: \qt{Last time, I assigned a homework about `your ideal future family', but [...] she just scribbled a bit without expressing any clear thoughts}. The therapists emphasized the importance of guiding clients in verbalizing their emotions alongside art-making. T5 mentioned that while visual art can help explore subconscious thoughts, verbalizing these feelings provides a cathartic outlet and helps clients externalize their emotions.

\subsubsection{\textbf{CH2}: Challenges in Customizing Therapy Homework} 

Our therapists demonstrated their practice of customizing homework assignments in art therapy. For instance, T2 and T5 mentioned tailoring homework tasks and specific instructions based on their practical experience and therapeutic techniques (e.g., cognitive-behavior therapy or mindfulness): \qt{If I suggest therapy homework that integrates mindfulness with art-making, I might ask the client to notice any changes in their breathing [during homework]}~(T4). T1 also adjusted homework tasks based on feedback from previous in-sessions.
However, the therapists noted that adapting structured instructions flexibly was difficult with current verbal or written formats, often leading to clients forgetting or abandoning their guidance or instructions~(\textbf{CH2-1}). Additionally, T3 and T4 observed that offering encouraging words and support during homework could boost motivation, but they found it challenging to provide personalized encouragement outside of in-session times~(\textbf{CH2-2}).


\subsubsection{\textbf{CH3}: Challenges in Tracking Therapy Homework History} 

The therapists confirmed that original homework data---such as the artworks, conversation records about clients' creative states, and details of the creative process---were essential for their assessments. They also encouraged clients to bring homework outcomes to the next session. For example, T1 and T3 prompted clients to share their current feelings and perspectives during one-on-one sessions, while T4 encouraged clients to engage in re-creation based on their homework.
However, therapists commonly expressed difficulty in tracking homework history, as they relied on clients to record and report their own progress~(\textbf{CH3-1}): \qt{The client drew [an artwork] two months ago. When you showed her the artwork, she often didn't remember what had happened at the time~(T3)}. Additionally, T1, T3, and T4 raised concerns that current practices might miss valuable data regarding clients' emotional or mental states at the time the homework was completed~(\textbf{CH3-2}).

\section{\name{} SYSTEM}
This section covers the design and implementation of \name{}. We first present the core design features developed based on our contextual understanding, followed by a typical usage scenario and the technical implementation details.

\begin{figure*}[tb]
  \centering
  \includegraphics[width=\linewidth]{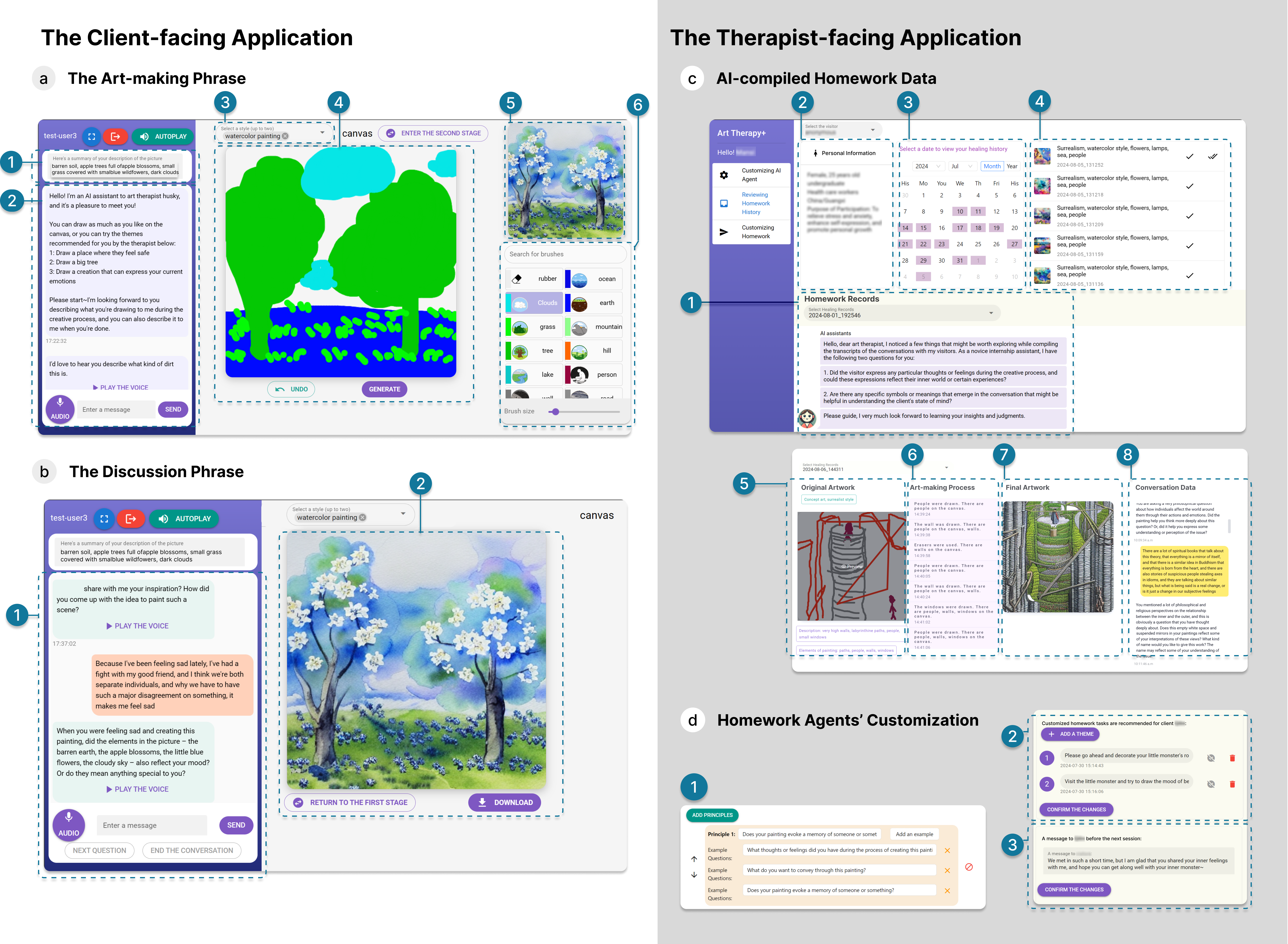}
  \vspace{-4mm}
  \caption{\name{} consists of the client-facing application and the therapist-facing application (is presented in translated English Version): (1) the client-facing application includes the Art-making Phrase interface and the Discussion Phrase interface; (2) the therapist-facing application includes Homework Agents' Customization interface and AI-compiled Homework Interface}
  \Description{Figure 2 presents two applications: the client-facing application and the therapist-facing application. The left side of the figure shows the client-facing application, while the right side displays the therapist-facing application.
  In the client-facing application, the upper section represents the "Art-Making Phase" interface. This interface includes a conversational agent, an art-making canvas for the user, a co-creative preview canvas, and tools that allow clients to select different brushes. The lower section is the "Discussion Phase" interface, which contains a human-AI co-creative artwork and a conversational agent that enables users to engage in dialogue with it.
  In the therapist-facing application, the upper section shows the "AI-Compiled Homework History" interface, where therapists can review the client's homework history. This interface includes the client's personal information, original creations, the creation process, the final artwork, conversation records, and two questions summarized by the AI assistant based on the dialogue history. The lower section displays the "Homework Agents Customization" interface, which allows therapists to set homework tasks, modify or add principles and sample questions for the conversational agents, and write personal messages for the client.}
  \label{fig:ui}
\end{figure*}

\begin{figure*}[tb]
  \centering
  \includegraphics[width=\linewidth]{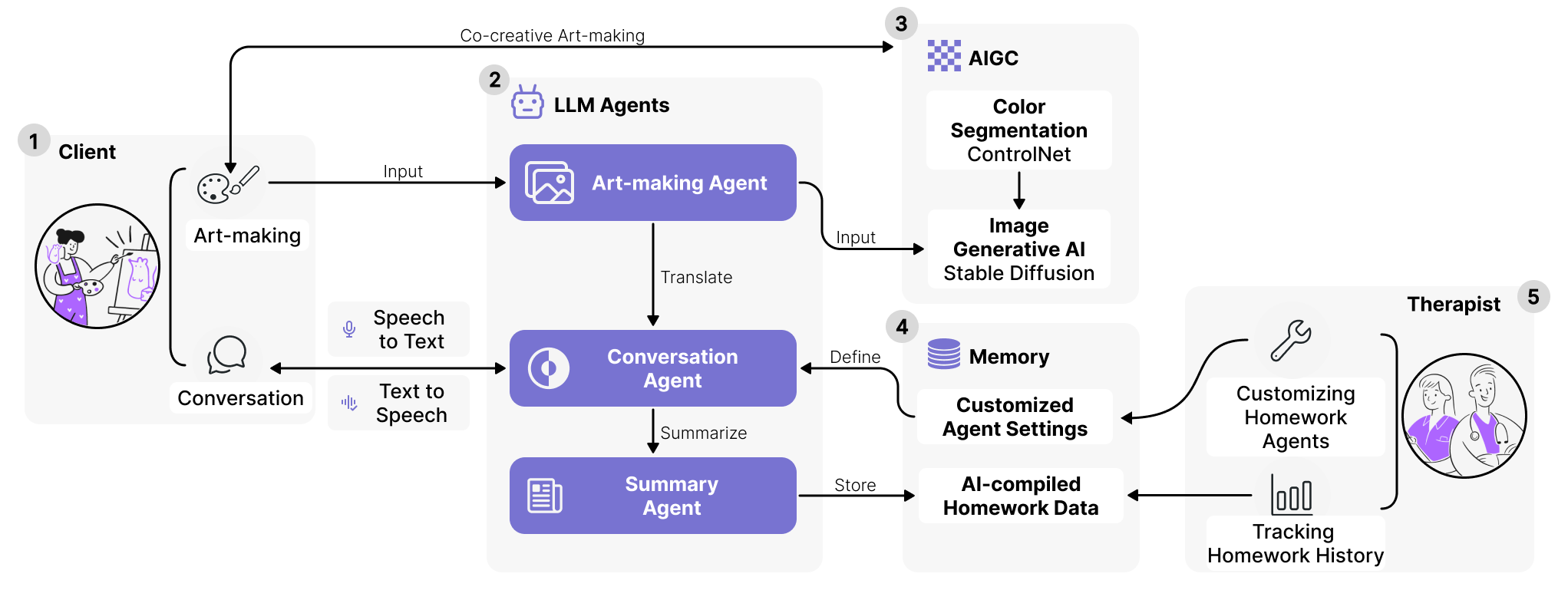}
  \vspace{-4mm}
  \caption{Overview of the \name{} system architecture}
  \Description{
Figure 4 illustrates the system architecture. On the left, the client-side focuses on two main functions: art-making and conversation. The art-making function is powered by ControlNet for color segmentation and the Stable Diffusion model. Additionally, an art-making agent helps clients describe their artwork, summarizing these descriptions into prompts, which are then fed into the Stable Diffusion model. The conversation function is built on large language model agents, and the system supports voice interactions between clients and agents through speech-to-text and text-to-speech APIs.
On the right, the therapist-side also has two main functions: customizing homework agents and tracking homework history. The customizing homework agents feature allows therapists to set personalized homework tasks and define the principles for the conversational agents. These customizations are integrated into the conversational agents. Additionally, summary agents are used to compile and summarize the client's homework data.}
  \label{fig:system}
\end{figure*}

\subsection{\name{} Core Design Features}
To understand how a human-AI system could support both clients' homework (\textbf{RQ1}) and therapist-client collaboration surrounding it (\textbf{RQ2}), we have designed \name{}, which consists of: 
(1) a client-facing application that combines human-AI co-creative art-making with conversational interaction to facilitate homework in daily settings, and (2) a therapist-facing application that offers AI-compiled homework history and customization of client homework agents for tailored guidance (see \autoref{fig:ui}). To address three key challenges we identified in CONTEXTUAL UNDERSTANDING, we collaborated closely with the therapists to develop the following core design features of \name{} (\textbf{DF1}–\textbf{DF3}):

\subsubsection{\textbf{DF1}: Combining Human-AI Co-creative Art-making with Conversational Interaction (Client-facing Application)}
To address \textbf{CH1}, the client-facing application leverages a human-AI co-creative canvas (\autoref{fig:ui} (a)) to lower the art-making threshold for clients, and a two-phase conversation workflow to provide clients with structured guidance in both the ``Art-making Phase'' and the ``Discussion Phase''. It features three panels: 
\begin{itemize}
    \item \textit{Human-AI Co-creative Canvas} including various AI brushes~(\autoref{fig:ui}~(a)-\circled{6}) that enable users to draw color-coded segments~(\autoref{fig:ui} (a)-\circled{4}) with each color mapped with a semantic concept, and translate these user-drawn forms into concrete objects, as previewed in~(\autoref{fig:ui}~(a)-\circled{5}). The rationale for enabling AI brushes is that clients can choose brushes representing a variety of semantic concepts, and they can create color-coded forms to directly control the shapes of the corresponding object. Their usage of the semantic brushes can also help therapists understand the conceptual elements of the artworks, which might project client's personality and emotions~\cite{malchiodi2007art}.
    \item \textit{Art-making Phase Conversation} is designed to encourage clients' self-expression by prompting them to verbally describe their artistic concepts as they create color-coded segments using the AI brush. These verbal descriptions are then summarized into text prompts, providing greater control over the generated images, as illustrated in \autoref{fig:ui} (a)-\circled{1} and \circled{2}.
    \item \textit{Discussion Phase Conversation} has been suggested by our therapists to facilitate deliberate self-exploration and reflection right after art-making through multi-round conversations, following our task instructions~(\autoref{fig:ui}~(b)-\circled{1}). The dialogue principles and example questions in the default task instructions came from the art therapy literature~\cite{buchalter2017250,buchalter2004practical,buchalter2009art} and were later customized by the art therapists.
\end{itemize}

\subsubsection{\textbf{DF2}: Supporting the Customization of Client Homework Agents (Therapist-facing Application)} 

To address \textbf{CH2}, customizing homework agents in the therapist-facing application includes three features:
\begin{itemize}
    \item \textit{Dialogue Customization} allows therapists to create, modify, or reorder dialogue principles and their example questions, controlling the dialogue flow of client's conversational agent in Discussion Phase. This enables the therapist to tailor the homework conversation (especially in the Discussion Phase) based on the therapist's professional belief and their understanding about the needs of a specific client~(\autoref{fig:ui}~(d)-\circled{1}).  
    \item \textit{Homework Task Customization} is designed to support therapists in tailoring homework tasks based on their understanding of the client~(\autoref{fig:ui}~(d)-\circled{2}), with these tasks then displayed in the client-facing application's dialog display area~(\autoref{fig:ui}~(a)-\circled{2}); 
    \item \textit{Opening Message Customization} was required by our therapists for tailoring the greeting message to a client, which are displayed in the dialog area of their client-facing application~(\autoref{fig:ui}~(d)-\circled{3}). These messages offer personalized encouragement and emotional support during art therapy homework.
\end{itemize}


\subsubsection{\textbf{DF3}: Enabling AI-compiled Homework History (Therapist-facing Application)}
To address \textbf{CH3}, a AI-compiled homework history interface is designed with three panels:
\begin{itemize}
    \item \textit{Homework Overview} including personal information, a visualization of usage and a short AI-compiled summary of each session~(\autoref{fig:ui}~(c)-\circled{2}, \circled{3} and \circled{4}); 
    \item \textit{Homework Records} show different outcomes of a homework session, including client-drawn color segments (\autoref{fig:ui}~(c)-\circled{5}), the client-AI co-created artwork~(\autoref{fig:ui}~(c)-\circled{7}), the art-making process~(\autoref{fig:ui}~(c)-\circled{6}), and the client-AI conversation history~(\autoref{fig:ui}~(c)-\circled{8}); 
    \item \textit{AI Assistant Summary} eases therapists' review of a client's homework session by summarizing based on the client's verbal inputs and art-making outcomes. It highlights content, experiences, feelings, and reflections expressed during the homework session (\autoref{fig:ui}(c)-\circled{1}). Following therapists' suggestions, it avoids interpretations and instead poses questions from a novice therapist's perspective, pointing out potential relevant aspects to aid the therapist's review.
    
\end{itemize}

\subsection{\name{} Usage Scenario}

Here we present a typical usage scenario of \name{}: Alice had been struggling with her relationship, feeling misunderstood by her partner, so she booked online art therapy sessions with Jessica, her therapist.
So she decided to book online art therapy activities with Jessica, her art therapist.
After their first session, Jessica used \name{}'s therapist-facing application.
Drawing from her experience, she incorporated dialogue principles and example questions into the system (\autoref{fig:ui}~(d)-\circled{1}). Jessica first added a dialogue principle of ``guiding users to describe the overall work'' and provided a few example questions for the conversational agent to reference: e.g., ``I would love to hear how you describe this work.'' Similarly, she added a few other principles and adjusted their order. The conversational agent can then follow these principles and examples in the given structure.
Further, Jessica also tailored homework tasks related to couple relationships (\autoref{fig:ui}~(d)-\circled{2}) and added a supportive message, ``Your sensitivity and ability to put yourself in others' shoes are truly a gift'' (\autoref{fig:ui}~(d)-\circled{3}), to offer encouragement during Alice's homework~(\textbf{DF2}).

Another day, after a quarrel with her boyfriend, Alice felt overwhelmed and turned to the homework Jessica had assigned using \name{}: drawing two plants, one representing herself and the other her partner. She opened the client-facing app and entered the ``Art-making Phase''~(\textbf{DF1}) on her tablet. She selected the \textit{Tree} brush from the toolbox~(\autoref{fig:ui}~(a)-\circled{6}) and began drawing a tree on the canvas, while the art-making agent prompted her to describe it through voice input~(\autoref{fig:ui}~(a)-\circled{2}). 
Alice described the tree as an apple tree full of blossoms, and the agent summarized text prompts about the current creation based on her input. 
She then added more objects like \textit{Soil}, \textit{Cloud}, and \textit{Grass} and completed her artwork. After selecting the \textit{Watercolor Painting}~(\autoref{fig:ui}~(a)-\circled{3}) style,
she clicked "Generate". The color segments drawn by the client, combined with the text prompts, served as inputs to produce a human-AI collaborative artwork that deeply resonated with her emotions.

Moving into the ``Discussion Phase'', the conversational agent guided Alice in reflecting on her artwork, asking questions like, ``Would you like to describe your tree?'' Alice shared her feelings and realized that she and her boyfriend were like two different plants—independent yet needing to understand each other.

Before their next session, Jessica reviewed Alice's homework history through the therapist-facing interface~(\textbf{DF3}). She examined the original and co-created artworks, along with the conversation records, gaining a deeper understanding of Alice's situation. Noticing that the AI assistant had prompted her to consider Alice's experience of arguing with her partner~(\autoref{fig:ui}~(c)-\circled{1}), Jessica decided to address these insights during their upcoming session. After revisiting Alice's homework sessions, Jessica decided to moved the principle ``naming the artwork'' from the second to the end of the dialogue. She believes this would help Alice better articulate her thoughts.

\subsection{\name{} System Implementation}
As \autoref{fig:system} shows, the \name{} system consists of the front-end of both client-facing application and therapist-facing application, as well as the back-end including LLM Agents module, AIGC module, and the memory module.



\subsubsection{Front-end}

The front-end of both client and therapist applications are web-based UIs built with Quasar Framework, which is based on Vue.js, and extends its capabilities by providing a rich UI component library, tools and cross-platform development support. 
For the conversational interaction of client-facing application, client messages are processed using OpenAI's TTS model\footnote{OpenAI TTS Models, https://platform.openai.com/docs/models/tts} to generate a corresponding voice message that autoplays by default. We also implemented speech to text function using the Voice Dictation (Streaming Version)\footnote{iFLYTEK Speech-to-Text, https://global.xfyun.cn/products/speech-to-text} service from iFlytek. For the therapist-facing application, all data is stored and transmitted in JSON format, interacting with the back-end server via HTTP requests, allowing therapists to select a client from a list of names, and view the session logs, assign customized homework, and modify the agent's guidelines.

\subsubsection{Back-end}

The backend is implemented using Flask, responsible for generating images, processing LLM workflows, and managing data. 
The image generation model pipeline uses \textit{ControlNetModel} and \textit{StableDiffusionControlNetPipeline} from huggingface's diffusers\cite{von-platen-etal-2022-diffusers}, and the base model is \textit{runwayml/stable-diffusion-v1-5}, the control net segmentation model is \textit{lllyasviel/control\_v11p\_sd15\_seg}\footnote{Controlnet - v1.1 - seg Version, https://huggingface.co/lllyasviel/control\_v11p\_sd15\_seg}. The system uses asynchronous, queue-based processing with multi-threading to efficiently handle multiple concurrent image generation requests. 

The backend for the LLM workflows consists of five prompted LLM Agents, all utilizing GLM-4\footnote{GLM-4 GitHub, https://github.com/THUDM/GLM-4}. 
The first LLM, refer to "Art-making Agent" of \autoref{fig:system}-\circled{2}, is utilized during the Art-making Phase of the client app, where it summarizes users' drawings and descriptions into prompts for the Stable Diffusion model. The second LLM, refer to "Conversation Agent" of \autoref{fig:system}-\circled{2}, is employed during the Discussion Phase, using a template that incorporates principles and customized questions from therapists. The third and fourth LLMs, refer to "Conversation Agent" of \autoref{fig:system}-\circled{2}, process data from the Art-making and Discussion Phases, respectively, and condense the information into short summaries for the therapist. The fifth LLM, also refer to "Conversation Agent" of \autoref{fig:system}-\circled{2}, integrates dialogue from both phases and generates three therapist-focused questions based on the conversation history. All the prompt are included in the supplementary materials of the paper.

These services communicate via RESTful APIs, the entire backend service has 15 APIs, ensuring the functionality and a smooth front-end and back-end interaction.
The back-end handles data storage and retrieval, managing user-generated content including AI-generated and original artwork, creative process data, homework dialogue data, color segmentation area data, and customized homework data. Images are stored in PNG format, while logs and settings are saved in JSON format.



\begin{table*}[tb]
\centering
\caption{Demographics of Participant Clients: Previous Art Therapy Sessions indicates the number of times the client has previously participated in art therapy; Familiarity with Traditional Drawing reflects the client's level of experience with traditional drawing techniques (0-not familiar; 1-very familiar); Familiarity with Digital Drawing reflects the client's level of experience with digital drawing techniques (0-not familiar; 1-very familiar); Participation Purposes reflects the reasons clients choose to engage in the activity.}
\vspace{-3mm}
\label{tab:clients}
\small
\resizebox{1\linewidth}{!}{
\begin{tabular}{cccccccccc}
\toprule
\textbf{ID} & \textbf{Gender} & \textbf{Age} & \textbf{Education} & \textbf{Region} & \parbox[t]{2.5cm}{\centering\textbf{Previous Art Therapy Sessions}} & \parbox[t]{3cm}{\centering\textbf{Familiarity with Traditional Drawing}} & \parbox[t]{2cm}{\centering\textbf{Familiarity with Digital Drawing}} & \parbox[t]{2cm}{\centering\textbf{Therapist Assignment}} & \parbox[t]{2.5cm}{\centering\textbf{Participation Purposes}} \\
\midrule
C1  & Female & 37  & Bachelor's & China/Shanghai & 0                            & 1                                   & 0.25  &T3 & Personal Growth                   \\
C2  & Female & 35  & Bachelor's & China/Shenzhen & 3                            & 0.5                                   & 0.5   &T3 & Career Development and Family                 \\
C3  & Female & 28  & Master's   & China/Hebei    & 2                            & 0.75                                  & 0.75   &T3  & Family and Emotional Management                \\
C4  & Female & 36  & Bachelor's & China/Beijing  & 10                           & 0.75                                   & 0   &T3  &Career Development                \\
C5  & Male   & 28  & Master's   & Germany       & 0                            & 1                                   & 0.75   &T3   &  Emotional Management and Personal Growth                       \\
C6  & Other  & 26  & Associate's & China/Heilongjiang & 1                            & 0.5                                   & 0.25  &T5  & Emotional Exploration and Intimate Relationships                           \\
C7  & Female & 23  & Master's   & China/Shanghai & 0                            & 1                                   & 1     &T5     &  Intimate Relationships                    \\
C8  & Female & 20  & Bachelor's & China/Shenzhen & 0                            & 0.5                                   & 0.5    &T5   &  Emotional Management and Intimate Relationships                       \\
C9  & Female & 25  & Bachelor's & China/Guangxi  & 4                            & 0                                   & 0.5    &T5    &  Self-Expression and Emotional Exploration                      \\
C10 & Male   & 23  & Master's   & China/Shenzhen & 0                            & 0.75                                   & 0.5   &T5   &             Self-Expression and Social Skills             \\
C11 & Female & 26  & Master's   & China/Hangzhou & 0                            & 0.5                                   & 0.25    &T4  &        Emotional Management, Social Skills and Intimate Relationships                 \\
C12 & Female & 26  & Master's   & China/Shanghai & 2                            & 0.75                                   & 0.5    &T4   &                   Stress Relieving and Intimate Relationships  \\
C13 & Female & 30  & Master's   & China/Dalian   & 0                            & 0.5                                   & 0.25   &T4    &             Family and Emotional Management            \\
C14 & Female & 19  & Bachelor's & China/Chongqing & 0                            & 0.25                                   & 0.25   &T4  &                Personal Growth and Self-Exploration           \\
C15 & Male   & 27  & Bachelor's & China/Beijing  & 0                            & 0.25                                  & 0.25   &T4    &                 Stress Relieving and Personal Growth        \\
C16 & Female & 22  & Bachelor's & China/Shandong & 0                            & 0.5                                   & 0.25   &T1     &              Emotional Management and Social Skills       \\
C17 & Male   & 38  & Master's   & China/Sichuan  & 0                            & 0.75                                   & 0.75   &T1     &                    Personal Growth      \\
C18 & Female & 40  & Master's   & China/Beijing  & 20                           & 1                                   & 0.75    &T1      &               Stress Relieving and Emotional Management          \\
C19 & Female & 28  & Bachelor's & China/Guangzhou & 0                            & 0.5                                   & 0   &T1       &                 Future Career Planning and Personal Growth      \\
C20 & Male   & 25  & Master's   & China/Guangzhou & 0                            & 1                                   & 1   &T1        &                    Academic Pressure Relieving   \\
C21 & Male   & 24  & Master's   & China/Hubei    & 0                            & 0                                   & 0   &T2        &                Childhood Family and Dreams Exploration  \\
C22 & Female & 24  & Master's   & China/Shenzhen & 0                            & 0.25                                   & 0.25    &T2  &                Emotional Management and Personal Growth     \\
C23 & Male   & 25  & Master's   & China/Zhejiang & 10                           & 0.5                                   & 0.5    &T2   &                  Emotional Development and Self-Expression        \\
C24 & Male & 55  & Bachelor's & Dubai& 0 & 0.5& 0.5&T2 &                           Emotional Management \\
\bottomrule

\end{tabular}}
\Description{The table 2 describes 24 participants in art therapy sessions. The participants are from diverse locations, including China (Shanghai, Shenzhen, Hebei, Beijing, Heilongjiang, Guangxi, Hangzhou, Chongqing, Shandong, Sichuan, Hubei, and Zhejiang), Germany, and Dubai. The ages range from 19 to 55 years old, with varying levels of education from associate degrees to master's degrees and bachelor's degrees. Their familiarity with traditional drawing techniques ranges from no familiarity to very familiar, while their familiarity with digital drawing techniques also varies across the spectrum. The participants have attended between 0 and 20 previous art therapy sessions and are assigned to different therapists identified by codes T1 to T5.Participation Purposes reflects the reasons clients choose to engage in the activity}
\end{table*}

\section{Field study}
Using \name{} as both a novel system to study and a research tool to study with, we aim to explore how a human-AI system support clients' art therapy homework in their daily settings (\textbf{RQ1}) and how such a system could mediate therapist-client collaboration surrounding art therapy homework (\textbf{RQ2}). To this end, we conducted a field deployment involving 24 recruited clients and five therapists over the course of one month.

\subsection{Participants and Study Procedure}
\subsubsection{Participants}

The five therapists who participated in the field evaluation were the same ones from our contextual study (see \autoref{tab:expert}). Each therapist was compensated at their regular hourly rate.
For client recruitment, we distributed digital flyers through social media platforms, describing the art therapy activities as an "online art therapy experience promoting self-exploration using a digital software." This aligns with the common goal of art therapy sessions, which are widely used to promote self-exploration for all clients, beyond treating mental illness~\cite{kahn1999art, riley2003family}.

Participants first completed a pre-questionnaire, which provided an overview of the activities and collected demographics, and prior experiences with art therapy experience and with digital drawing---to ensure that we include both novices and experienced user---and their personal goals for participation. 
The therapists guided the recruitment and screening of the the clients, and included individuals seeking for reducing stress, fostering personal growth, enhancing emotional regulation, and strengthening social skills. The therapists excluded individuals with serious mental health conditions to minimize ethical risks.

In total, 27 clients began using \name{}, but 3 withdrew early due to scheduling conflicts. The final group of 24 clients (C1-C24; 8 self-identified males, 15 self-identified females, 1 identifying as other; aged 19-55) completed the study (client demographics are detailed in the~\autoref{tab:clients}). Clients who completed the full process were compensated with \$37, others were compensated with a prorated fee.
Our study protocol was approved by the institutional research ethics board, and all participant names in this paper have been changed to pseudonyms. Participants reviewed and signed informed consent forms before taking part, acknowledging their understanding of the study.

\subsubsection{Procedures}

Clients were distributed in coordination with the five therapists, as shown in \autoref{tab:expert}. T2 was assigned four clients, while the other therapists each had five clients. The field study consisted of two main activities: (1) three online in-session activities, where clients had one-on-one conversations and collaborated with the therapist, and (2) unstructured between-session activities, where clients practiced therapy homework using \name{} following the therapist’s recommendations.
Before the study, we held online introductory sessions to familiarize the clients with \name{}, and provided both demonstrations and hands-on exploration on their preferred devices. Similarly, we offered online training for therapists on customizing and reviewing homework, while allowing them to explore both the therapist-facing and client-facing applications. After the session, clients were encouraged to regularly explore \name{}.
Two weeks into the study, we scheduled weekly one-on-one online sessions between therapists and clients, each lasting approximately 60 minutes. Therapists were encouraged to review the clients' homework history using \autoref{fig:ui}(c) before each session. During the online session, therapists used this data to inform their practices without interrupting the flow of therapy. We encouraged clients in advance, to create artworks during the Art-making Phase~(\autoref{fig:qual_results}(a)), sharing screens and discussing their creations with the therapist, but did not interfere with the therapeutic process.


At the end of each session, therapists recommended homework tasks based on insights gained during the conversation. After the session, therapists might customize homework agents, including customizing conversational principles, assigning homework tasks, and providing personal messages through \autoref{fig:ui}~(d). Clients could then either complete the assigned homework or engage in self-exploration using \name{} between sessions.

\subsection{Data Gathering Methods} 

For between-sessions, we stored all homework-related data in a database, including artwork, dialogue, usage logs, as well as information on homework customization such as conversational principles, tasks, and personal messages.
We encouraged participants to use personal messaging (WeChat) to share pictures and comments about on-the-spot experience and feelings after homework with \name{} to compensate for semi-structured interviews.
During online sessions, we recorded audio and video. 
The researchers did not observe the therapy session in live, but reviewed post hoc, as the therapists believed a third party's presence could affect a client's emotional expression and the therapist-client dynamic.
After each session, we conducted a brief 5-minute interview with the therapists to gather their insights and feelings.

Upon the completion of the final one-on-one sessions, we conducted 30-minute semi-structured interviews with both therapists and clients. These interviews aimed to explore how \name{} supported art therapy homework in clients' daily lives (\textbf{RQ1}) and how therapists and clients collaborated surrounding art therapy homework (\textbf{RQ2}). We used feedback and homework outcomes from the trial period to ask targeted questions about their practices.
With participants' consent, we recorded and transcribed the brief 5-minute interviews and the 30-minute interviews for thematic analysis~\cite{braun2006using}. This analysis also included the personal messages shared by the participants about their on-the-spot experiences.
Two researchers then engaged in inductive coding, annotating transcripts to identify relevant quotes, key concepts, and patterns. They developed a detailed coding scheme through regular discussions, grouping quotes into a hierarchical structure of themes and sub-themes. Exemplar quotes were selected to represent each theme. We also used homework history (e.g., images or conversation data) and customization data (e.g., homework dialogue principle data) as evidences or examples to back up the findings in our thematic analysis.

\begin{figure*}[tb]
  \centering
  \includegraphics[width=\linewidth]{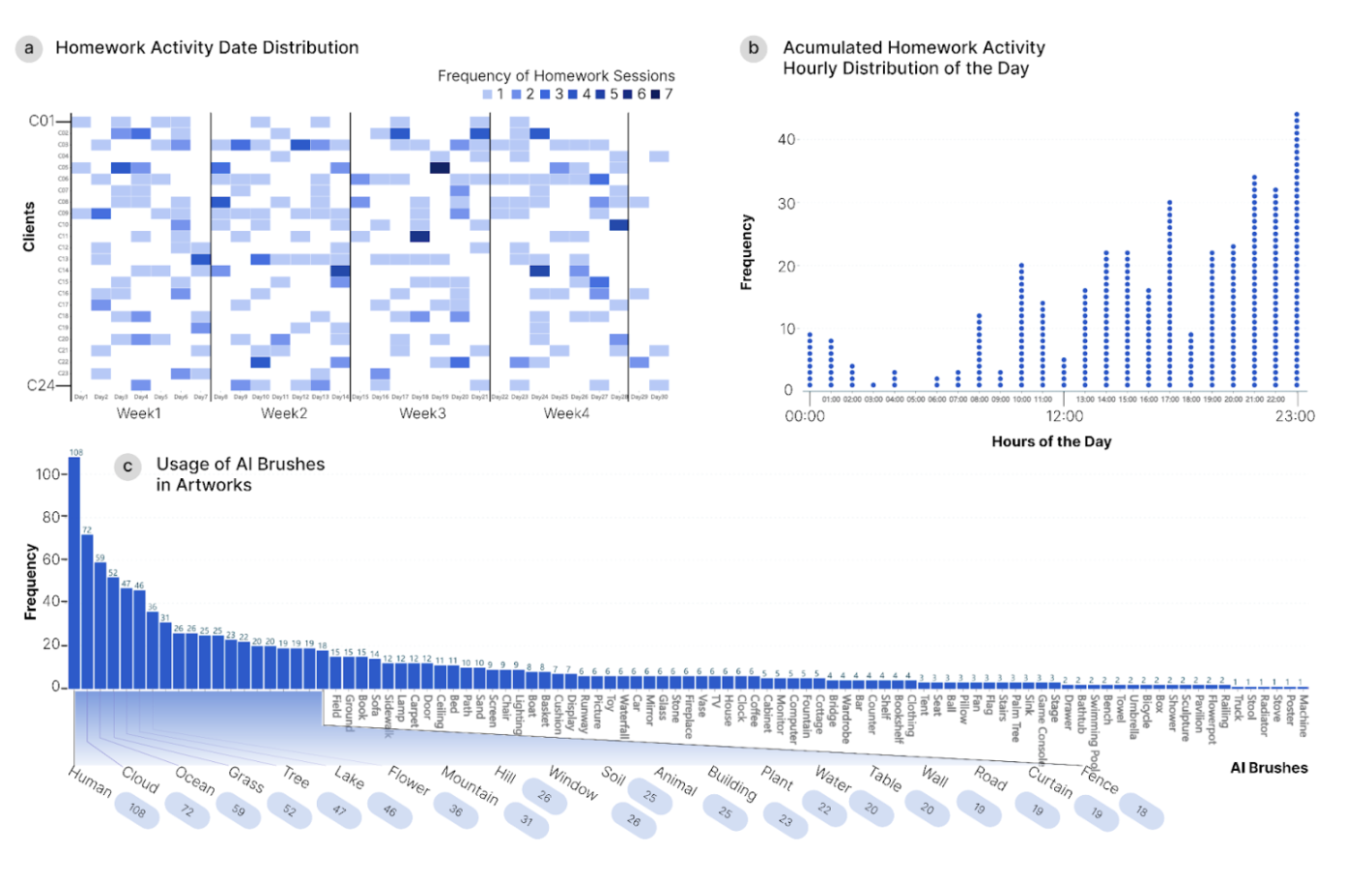}
  \vspace{-7mm}
  \caption{Overview of The Homework Engagement of Clients with \name{}: (a) Homework Activity Date Distribution; (b) Accumulated Homework Activity Hourly Distribution of the Day; (c) Usage of AI Brushes in Artworks; 
  }
  \Description{Figure 5 contains three sub-figures. Figure 5a shows the Homework Activity Date Distribution for 24 clients over a four-week period, using seven different shades of purple to represent varying levels of participation in the homework sessions. Figure 5b illustrates the frequency of AI brush usage during clients' homework art-making, with the top 20 most frequently used brushes highlighted in larger font. Figure 5c depicts the distribution of homework sessions across different times of the day, revealing that clients tend to engage in homework sessions more frequently in the afternoon and evening.}
  \label{fig:quan_results}
\end{figure*}

\begin{figure*}[tb]
  \centering
  \includegraphics[width=\linewidth]{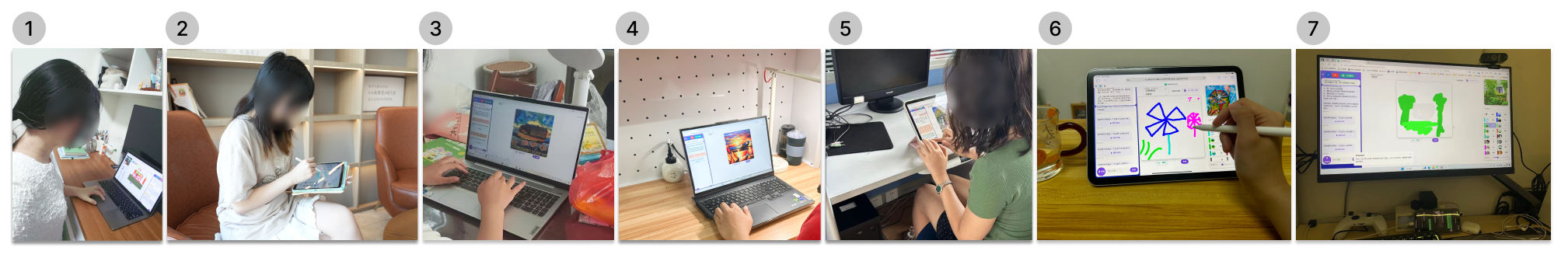}
  \vspace{-7mm}
  \caption{Photos provided by clients showing their typical settings while using \name{}: (1) C2 in her bedroom; (2) C7 in a library; (3) C16 at home; (4) C10 in his dormitory; (5) C22 in her office; (6) C9 at home; (7) C21 at home. (Note: Clients were invited to pose for these photos.)}
  \Description{
Figure 6 shows seven images: (1) C2 used a laptop in her bedroom, (2) C7 used a tablet in the library, (3) C16 used a laptop at home, (4) C10 used a laptop in his dormitory, (5) C22 used a tablet in her office, (6) C9 used a tablet at home, and (7) C21 used a PC at home.}
  \label{fig:context}
\end{figure*}

\section{Findings-RQ1: How \name{} supported clients' art therapy homework in their daily settings}
This section reports how the clients, over the period of one month, utilized the \name{} system for art therapy homework in thei natrual settings.

\subsection{Overview of Usage}

We provide an overview of client engagement with \name{} throughout the 30-day period. 24 clients completed 354 homework sessions, with 266 including both art-making and discussion phases and 88 only art-making phases. They spent totally 151 hours---66.4 hours on art-making and 84.6 hours on discussion phases. While interacting with the homework agent, they produced 3,462 messages---772 during art-making and 2,690 during discussion phases. On average, each client used the system 14.75 times. Art-making phases averagely took about 11.3 minutes, and discussion phases around 19 minutes.

\autoref{fig:quan_results} (a) shows each client's usage distribution over the 30 days. 
\autoref{fig:quan_results} (b) depicts the distribution of all clients' usage over different hours of the day, showing that clients more often completed homework in the afternoon and evening (13:00–23:00). 
\autoref{fig:quan_results} (c) lists the usage frequency of different AI brush objects in the artworks: \textit{Human} (108), \textit{Cloud} (72), \textit{Ocean} (59), \textit{Grass} (52), \textit{Tree} (47), \textit{Lake} (47), \textit{Flower} (47), etc. The frequent use of the \textit{Human} brush may reflect the significance of human figure drawings, often used by therapists to understand clients' personalities, developmental stages, and emotional projections~\cite{malchiodi2007art}.

\subsection{Contextualizing Clients' Homework Usage in Various Natural Settings}
This section contextualizes how clients used \name{} to engage in therapy homework in their daily settings~(\autoref{fig:context}). 
Our clients used different digital devices to complete their therapeutic work, such as laptop~(C1-C7, C10, C14, C15, C17, C18, C20-C22, C24), desktop computer~(C12, C18, C19, C21) and tablet~(C2, C4, C7-C11, C13, C22, C23). 
Also, our clients completed their therapy homework in various locations. All clients have used \name{} in their homes. Other common locations included dormitories (C10, C11), libraries (C7, C9, C10), cafes (C22), high-speed trains (C22), internet cafes (C18), hotels (C18), offices (C12, C17, C22), and even hospitals (C3).

\begin{figure*}[tb]
  \centering
  \includegraphics[width=\linewidth]{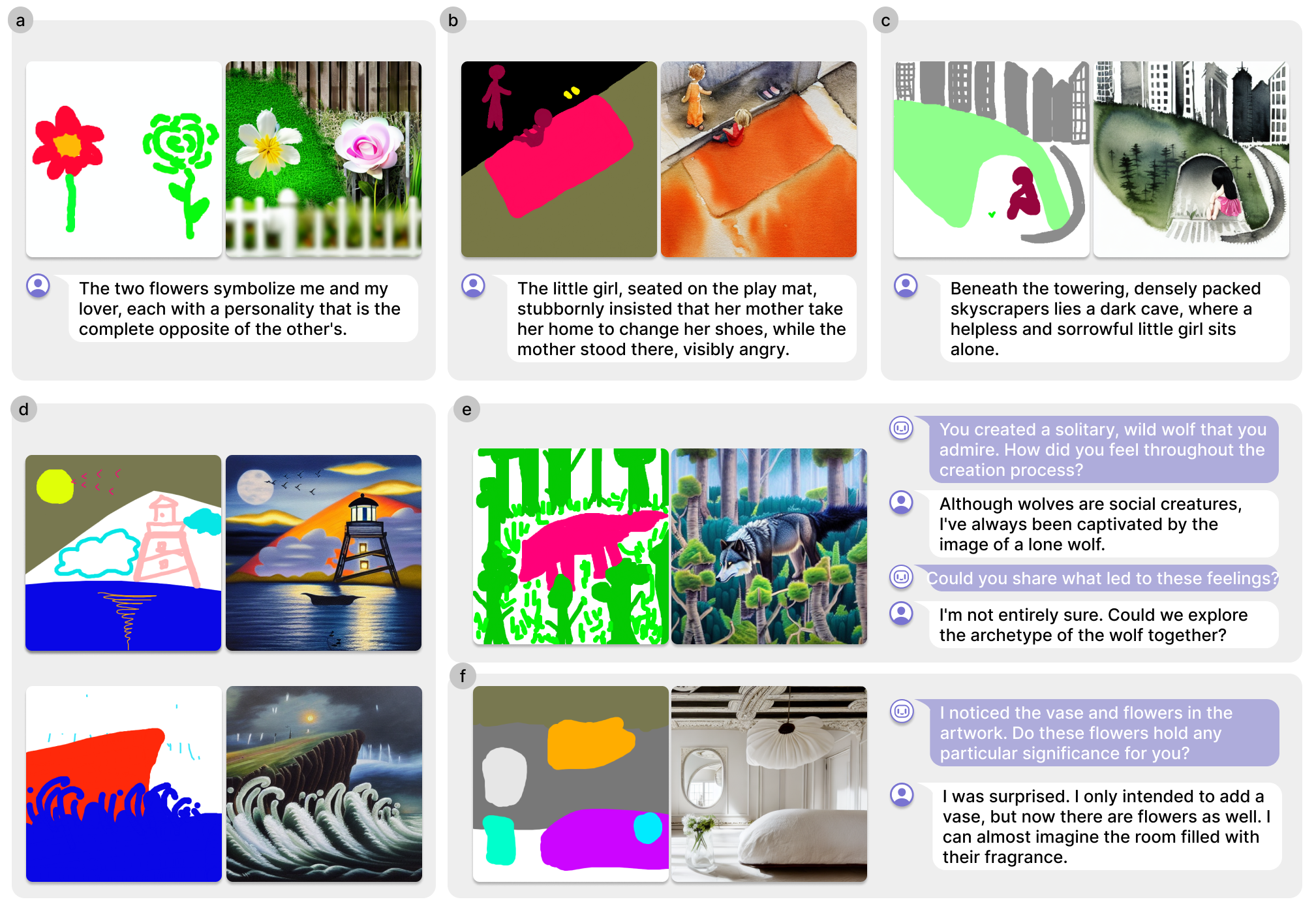}
  \vspace{-7mm}
  \caption{Outcomes of the Clients' Art Therapy Homework Supported by \name{}. 
  }
  \Description{Figure 7 illustrates the Outcomes of the Clients’ Art Therapy Homework Supported by TherAIssist. Figure 7(a) shows two flowers of different colors, which, through AI-assisted dialogue, were revealed to symbolize the client and her boyfriend.Figure 7(b) presents an artwork featuring two figures, a carpet, and a floor. In conversation with the AI conversational agent, the client externalized her experience of arguing with her daughter through both verbal expression and art-making.Figure 7(c) depicts a cave in front of towering skyscrapers, with a little girl and a small patch of grass inside. Through interactions with the AI, it became clear that the client used verbal expression and art-making to externalize her feelings of loneliness and helplessness.Figure 7(d) contrasts two images, both using the sea to represent the client’s emotions. In the first image, a calm sea, gentle moonlight, and a lighthouse convey a sense of peace and warmth. In the second image, chaotic waves symbolize the client’s inner anger.Figure 7(e) shows a wolf in a forest. Through dialogue with the conversational agent, the client and the conversational agent explored the meaning of the wolf step by step. Figure 7(f) displays a family bedroom created using different brushes, including elements like a vase and a bed. During the conversation with the AI, the client expressed surprise at the inclusion of a bouquet in the vase, an unexpected detail that delighted him.}
  \label{fig:qual_results}
\end{figure*}

The clients reported rich and vivid examples contextualizing how they used \name{} in daily settings and how their usage was shaped by the material and social surroundings, e.g., C8 used \textit{Flower} brush on her tablet to create artwork after an argument with her boyfriend (see \autoref{fig:qual_results} (a)). She told the conversation agent that the two flowers represented their personalities: \qt{I drew plants for each of us, noticing their differences. Believing everyone has a unique personality, I didn't dwell on the minor variations after finishing the drawing}.
Dream exploration is part of therapy homework, helping clients use dream materials to explore unconscious thoughts through creative expression~\cite{grotstein2009dreaming,freeman2002dreams}.
Nine clients used \name{} to explore and interpret their everyday dreams through artwork and conversation, e.g., C21 recreated a dream of becoming a teacher in a skyscraper office, expressing sadness to the AI about time passing since graduation.
These examples show that \name{} extends art therapy into clients' daily lives, helping them apply insights and changes from therapy and reflect on themselves through real-life experiences~\cite{kazantzis2007handbook}.
 
Three clients explicitly noted that the system lowered the threshold for accessing art therapy homework by allowing anytime access and overcoming time and space limitations, e.g., \qt{last time, when I got angry at my child, I immediately reached for my tablet [to use \name{}], because it's the most convenient method for me~(C2)}. 
T4 added that: \qt{Art therapy homework typically requires clients to sit down with a desk. [With \name{}] clients can use a tablet for therapy homework on the way, reducing space and setting constraints}. 
Meanwhile, seven clients noted that \name{} also seemed to lower the threshold for art-making. 
As noted by C2, \qt{I didn't need drawing skills. If I wanted to draw a streetlamp, I only needed to sketch its general shape, and I could describe the finer details using language}. 
C23 explained that it might help focus on the process of creation itself: \qt{I found this drawing method easy—just fill in the spaces, like a digital coloring book. It helped me focus more on the ideas}. 
T4 explained that \name{} might makes the process \qt{effective and fun} due to its low skill threshold, which could help clients \qt{focus more on the creative process without being stressed.}

\subsection{How Conversation and Art-making were Combined in Therapy Homework}
This section exemplifies how the clients---while interacting with \name{}---engaged in both verbal expression and art-making that granted therapeutic meanings.

\subsubsection{\textbf{Clients' verbalization of feelings and experiences during art-making}}
In the ``Art-making Phase'', we found that the participants tended to verbalized their experiences and emotions related to their artworks in two aspects:

\textbf{Clients verbalized emotional feelings during art-making.} 
In the ``Art-making Phase'', we found that the clients verbalized emotional expressions during their art-making process through the conversational agent.
In C2's therapy homework, she created an artwork featuring a \textit{Cave}, a \textit{Girl}, a \textit{City}, and a blade of \textit{Grass}, verbalizing the helpless and sorrowful girl in the dark cave, overshadowed by the dense and cold skyscrapers of the city through the conversational agent~(see \autoref{fig:qual_results} (c)).
T3 remarked that: \qt{her artwork and verbal descriptions might express a deep sense of loneliness, along with a yearning for prosperity intertwined with feelings of confusion}.
Further, clients verbalized their emotional expressions integrated into the non-verbal expressions behind the brush objects.
For instance, both C22 and C18 utilized the same brush object, Sea, to depict different oceanic forms in their artwork (see \autoref{fig:qual_results} (d)). C22 portrayed a calm sea surface to evoke a sense of leisure and tranquility, whereas C18 expressed suppressed anger through the depiction of turbulent waves: \qt{the ocean object feels quite oppressive. Coupled with her verbal description, it suggests that she might be experiencing significant pressure and underlying anger~(T1)}.

\textbf{Clients articulated narratives and personal experiences about their artwork.}
In the ``Art-making Phase'', we found that our clients verbalized the articulation of personal narratives and experiences behind an artwork.
For instance, C2 utilized various brush objects, including \textit{People}, \textit{Floor}, \textit{Carpet} and \textit{Shoes}, to create an artwork.
She sought to externalize her artwork, viewing her experience of an angry mother stands while her daughter sits on the carpet and has a heated argument with her~(see \autoref{fig:qual_results}~(b)).
T3 interpreted that: \qt{there was a noticeable sense of inner conflict in her experiences with raising her children}.
Based on relevant theories~\cite{carlson1997using,harpaz2014narrative}, this collaboration of AI-infused art-making and conversational agents can identify the ``signs'' embedded in the artwork through verbalization, while also guiding and complementing the narrative behind them.

\subsubsection{\textbf{Clients' meaning-making with the homework conversation agent after art-making}}
In the ``Discussion Phase'', we found that combining human-AI co-creative artwork and the conversational agent can facilitate mean-making, in twofold meanings:

\textbf{Clients' self-exploration behind the artwork within the multi-round human-AI dialogue.} 
We found that the multi-round dialogue in ``Discussion Phase'' encouraged clients' open-ended self-exploration and helped them discover new meanings within their co-creative artworks. 
For example, C6 utilized the \textit{Tree}, \textit{Grass}, and \textit{Animal} brush objects to created a vibrant, towering green forest inhabited by an adult blue wolf, while the conversational agent facilitated exploration of the qualities admired in the solitary wolf through multi-round structured questioning~(see \autoref{fig:qual_results} (e)).
Further, C9 employed the \textit{Flower} to create a flower bush: \qt{...I drew flowers to express my feelings about an activity. When I asked AI about their meaning, it prompted me to reflect on times I had 'blossomed.' At first, I regretted not having many such moments, but after several rounds of conversation, I realized they are beautiful and valuable, even if rare, and gained clarity about my true emotions}.
T5 explained that the co-creative meaning-making can \qt{assist clients in transforming their subconscious creations into narratives with personal meanings}.
Compared to the monologic forms of traditional therapy homework, multi-round dialogue avoids a single, closed perspective, allowing diverse perspectives, encouraging clients to engage in meaning-making and promote their understanding of the images~\cite{bacstemur2021integration,pare2004willow}.

\textbf{Clients' interpretation triggered by the unexpected details from human-AI co-created artwork.} 
Ten clients noted that unexpected details in AI-generated images sparked creative interpretations and enriched their stories verbally. 
For example, C23 created a simple room with a luxurious bed and vase, but the AI added a flower bouquet, prompting C23's comment during the Discussion Phase: \qt{I was pleasantly surprised, as I had only wanted a vase, but with the flowers added by the AI, it looks beautiful, and I can already imagine the room filled with their fragrance}~(see \autoref{fig:qual_results}~(f)).
Also, C5 used the \textit{Clock} and \textit{People} to conveys the guilt and sorrow of losing loved ones and the ache of their absence as time flows: \qt{I initially planned to create a space with clocks to express guilt about time, but the AI generated flower baskets instead...Perhaps these flower baskets symbolize the outcomes brought by time}.
T4 commented that the unexpected details might help the clients broaden their horizons: \qt{The generative AI feels more like a `familiar bystander', offering its own perspective. The unexpected details might can spark new ideas or provide a fresh lens, which may inspire clients to find a direction to move forward}.



\subsubsection{\textbf{Risks and Concerns}}
Although combining conversation and art-making in therapy homework offers many benefits, there are a few risks and concerns in three key areas:

\textbf{AI-infused art-making lacks multisensory experiences.} 
C22 and C23 noted that their AI-infused art-making process lacked the multi-sensory experience often associated with traditional materials.
As put by C23, \qt{The sound of traditional pens on paper is irreplaceable by generative AI. My paper artwork has depth—controlled pen pressure and watercolor concentration create rich textures, unlike AI’s one-dimensional results}. 
C22 also mentioned that: \qt{I can even smell traditional art-making materials, and I truly enjoy listening to the sounds they make on paper}.
Thus, T2 explained that the limitation of generative AI: \qt{The most remarkable aspect of generative AI as a material is symbolic, not sensory or physically tangible}. 
Considering the sensory aspect of art-making, it may promote clients' self-soothing and the expression of their inner sensations~\cite{lusebrink1990imagery}.

\textbf{Uncertainty in AI-infused art-making.} 
C9, C10, C11, C15, C20 and C23 mentioned that generative AI sometimes doesn't fully generate the images that the client expect. 
For instance, C20 shared feeling disappointed when the AI sometimes didn't generate what he wanted: \qt{I want to create a scene where clouds are raining on a perfect sea, but the AI cannot generate that perfect sea}. 
Also, C15 noted that the generative AI sometimes doesn't fully generate highly abstract images: \qt{
The morning after the Olympic opening ceremony, I tried creating an abstract LGBTQ+ multicultural artwork with postmodern symbols and a central white light beam using \name{}. When the AI couldn't generate it, I described the concepts instead verbally}.
T2 expressed concerns that when the images didn't meet expectations, it might lead to feelings of disappointment: \qt{When using paper and pencils, clients rarely blamed the materials for imperfect drawings...but, if the AI doesn't perform as expected, frustration is often directed at it, especially when unexpected symbols or triggers appear, heightening disappointment}.

\textbf{The LLM agent cannot capture the fine details of artworks.} 
We found that C5 and C21 mentioned that the conversational agent sometimes had difficulty fully understanding all the details of the images, e.g., \qt{the AI may miss details in my artworks, like whether the teacher in my office is standing or sitting, making it hard to ask specific questions, which can leave me feeling disappointed~(C21)}. 
T2 and T3 noted that the conversational agent may lack emotional resonance and intuition, potentially missing important details in a client's artwork and causing disappointment: \qt{C1 drew a large stone in her artwork, which seemed significant to me. I would ask her what the stone represents, but the conversational agent may not capture such details~(T3)}. 
But, T4 also expressed concern that: \qt{if it can understand the details and give some interpretations, it may mislead the clients' emotions and self-cognition}.



\section{Findings-RQ2: How \name{} Mediated Therapist-Client Collaboration surrounding art therapy homework}
This section outlines \name{}'s role in facilitating asynchronous therapist-client collaboration for art therapy homework, covering therapist customization of homework agents and the tracking and use of clients' homework history.

\subsection{How therapists Customized Homework Agents to Support Client Homework Asynchronously}
In this section, we concretely examine how therapists tailored the homework agents to support clients' therapy homework. 

\subsubsection{\textbf{Therapists used customization to provide structured guidance to clients}}
Our therapists tended to provide structured guidance through infusing their practical experience and professional beliefs into the homework agents in two aspects:

\textbf{Therapists provided structured guidance through customized dialogues.} 
First, T1, T4, and T5 placed the principle of naming the artwork at the end of all dialogue principles based on her own work experience,e.g., \qt{if clients wait until they finished discussing artwork before naming it, it can help them understand the artwork more deeply~(T1)}. 
Furthermore, T2, T4, and T5 revised the principles and example questions for the conversational agent. 
For example, T2 rephrased all of the example questions and added extra example questions from the dialogue principles based on her professional beliefs~(e.g., adding \qt{If this artwork could communicate with you, what part of it would it say to you}). 
C23 mentioned that it can encourage him to reflect more deeply: \qt{T2 hoped I could have an external dialogue with elements of my creation. Today, I drew a bed, and the conversational agent asked what the bed would say to me, allowing me to explore its deeper meaning}.
Also, T3 added a dialogue principle at the beginning: \qt{determine whether the artwork aligns with your expectations}. T3 also explained her reasons: \qt{I would like to understand to what extent the artwork meets their expectations, and whether different scores reflecting varying expectations lead to different thoughts}.
Therefore, customizing structured guidance could help guide the client in providing the information the therapist was seeking: \qt{Because I asked questions about aspects, clients may be more inclined to complete the homework about the aspects~(T4)}.
Also, customizing structured guidance through the conversational agent might promote a more flexible and open perspective on facilitating exploration: \qt{Previously, I gave clients fixed, printed questions as reflective prompts for the homework, but the customized conversational agent offers a more flexible, chat-like experience~(T4)}.

\textbf{Therapists provided structured guidance through personalized homework tasks.} 
All therapists tended to provide structured guidance through tailoring different types of homework tasks based on their theoretical orientation and practice style they followed. 
T3 and T5 incorporated positive psychology techniques to provide different types of homework tasks based on the various stages of therapy, e.g., \qt{in the first week, I suggested that C6 used \name{} to draw her inner emotional `monsters' as a way to express his feelings. In the second week, I encouraged him to draw his positive resources using \name{}~(T5)}. 
Further, T1, T2, and T4 would like to integrate meditation techniques into the art therapy homework assignments. 
For instance, T2 assigned C23 a structured homework task: \qt{close your eyes, focus on your breathing, meditate, and imagine your relationship with your boyfriend. What visual image would you use to express it?}, T2 explained, \qt{at first, I guided them to experience mediation. As they drew, it helped clarify and deepen their understanding of their feelings}.
The structured guidance through tailoring homework tasks might assist clients in clarifying specific task goals and reinforcing insights from therapy sessions~\cite{santisteban2003efficacy, tompkins2004using}.

\subsubsection{\textbf{Therapists used customization to retain personal touch and offer emotional support}}
We found that therapists customized homework agents to provide emotional support and encouragement for clients, in two aspects:

\textbf{Therapists customized homework conversation agents to enhance personal touch and foster trust.}
T2 and T4 noted that they modified the phrasing of example questions or conversational principles of the conversational agent to enhance the personal touch and foster trust between the client and therapist. 
For example, T2 adjusted the example question to make it more conversational and human-like, following the principle of guiding the user to describe their artwork (e.g., \qt{Would you like to share your thoughts? If not, I can quietly sit with you. But if you'd like to talk, I'm here and ready to listen}): \qt{I hope the AI's companionship might be conveyed in a more gentle manner}.
Also, T4 revised the conversational principle for asking clients about their feelings toward the artwork (e.g., \qt{If the user is willing, gently explore the reasons behind their feelings. They can choose not to respond if they're not ready}): \qt{AI conversational principles, based on theory, can incorporate a personal touch, helping to build trust and rapport with the client}.

\textbf{Therapists customized opening messages to deliver personalized emotional support and encouragement.} 
T2, T3, T4 and T5 demonstrated that they tended to infuse their emotional support and encouragement into therapy homework through delivering their opening messages in \name{}.
For example, T2 left an opening message for C23 (e.g., \qt{Find a space, infuse it with vitality, and let it continue to thrive}): \qt{The message symbolizes that the connection between the therapist and client persists beyond the therapy sessions}. 
Further, T3 left opening messages~(e.g., \qt{Focus on your needs in negative emotions, recognize your strengths, and gain insights into your inner feelings}) on \name{} based on discussions with C2: \qt{I hope to give them some emotional support or encouragement when they are doing the homework, so I tried to write this based on the process I communicate with them}.
Our therapists incorporated their personal touch to the therapy homework through our agents, which can promote more self-disclosure among clients during doing therapy homework: \qt{When I read the message, I feel a deep sense of warmth and care...It seems to encourage me to open my heart to create~(C9)}.
Tailoring homework by incorporating expressions of empathy and positive regard (such as encouragement and affirmation) might significantly contribute to the establishment and maintenance of a strong therapeutic relationship~\cite{cronin2015integrating}.

\subsubsection{\textbf{Risks and concerns}}
While customizing homework agents offers several benefits, there are a few risks and concerns in two main areas:

\textbf{Limited options for current customization.}
First, T1 and T5 noted that the lack of diverse AI conversational agent voices may cause discomfort among clients: \qt{Having a couple of voice choices could make the interaction more personal. Some clients may prefer talking to an agent of a certain gender, which could facilitate approachability and connection~(T1)}. 
Further, T3 and T4 mentioned that AI-infused art-making is primarily limited to drawing, highlighting the need to incorporate and customize other forms such as collaging, photography, and sculpture. 
T4 mentioned that diverse forms of art-making have varying effects on different clients: \qt{In my practice, we sometimes used photography, popular among seniors. Limited art forms may restrict their self-expression across various media}.

\textbf{LLM agents' rigidity and lack of social intuition in conversation} 
T1 and T2 highlighted that tailored conversational agents can occasionally result in rigid questioning and insufficient social intuition, e.g., C21 noted that: \qt{Although I shared many feelings and experiences, the responses often felt formulaic, lacking emotional depth, and were driven by rigid questions that kept asking about the reasons behind everything.}
T1 worries that: \qt{Such structured questioning might make clients feel constrained, unable to freely express their thoughts or emotions, which might affect therapeutic outcomes}.
Further, T3 commented: \qt{the conversational agent sometimes missed cues to stop asking questions when clients were reluctant. How can we signal AI to cease inquiries? Unlike therapists, who can read facial expressions and tone to decide when to halt, the conversational agent might lack this non-verbal intuition}.



\subsection{How AI Compiled Homework History Was Used by Therapists as Practical Resources}
Our design feature, \textbf{DF3}, aims to track the clients' homework history through providing the artworks, the art-making process, the dialogue history data, and the AI summary.
In this section, we describe how therapists tracked and utilized AI-complied therapy homework history~(\textbf{RQ2}). 


\subsubsection{\textbf{Therapists learned about clients' past experiences and characteristics from homework history}}
All therapists explained that they employed the homework history to deepen understanding of the clients' past experiences and characteristics, in two key aspects:

\textbf{Therapists learned about clients from their artworks.} 
T1-T4 mentioned that they learned about clients' past experiences and characteristics from the co-creative process and AI-generated artworks.
First, T4 reflected that she gained a deeper understanding of clients' current life and attitudes through the pattern of the human-AI co-creative process: \qt{
Some frequently switched styles, some gave up after a few tries, while others persisted in regenerating images or accepted the first result as perfect. Their reactions reflected broader life attitudes, and I strove to maintain a holistic understanding of each client}.
In addition, therapists understood the characteristics of clients through co-creative artworks.
T3 reviewed C5's creation of a horse with long horns placed inside a vase: \qt{He often placed conflicting elements within the same artwork, which might reflect his own inner conflicts and contradictory characteristics}.
Finally, T2 highlighted the recurring brushes in clients' artwork summarized by AI: \qt{AI reminded me that C22 recently incorporated velvet into her artwork, which prompted me to pay closer attention to the element}.
T2 explained that she used the brush elements as a crucial tool to explore clients' unconscious mind: \qt{In Jungian terms, what does `summer' symbolize in the collective unconscious? The brush elements clients often use may unconsciously reflect their emotional responses and the processing of these experiences}.
Thus, T4 explained that : \qt{The artworks history, referred to as a `behavior pattern', can reveal consistent trends over time. I believe these patterns may have led me to form some hypotheses about the clients}.

\textbf{Therapists learned about clients from homework dialogue history.} 
All therapists demonstrated that they learned about clients' experiences and characteristics from homework dialogue history data.
For example, T4 learned about how the client processes the current emotional state via dialogue data: \qt{One of C11's homework reflects a quarrel with her boyfriend that left her in a bad mood. She often used the AI system when feeling emotionally unsettled, which can help her calm down...}.
It can help therapists gain a more accurate understanding of how clients are currently managing their emotions: \qt{In art therapy, I focus on how clients manage emotions, especially intense anger. For example, one client created a piece of art in anger, but by our next session, his anger had subsided. This system provides a valuable, real-time record for clients who are experiencing high levels of emotional fluctuations~(T3)}.
Also, the interaction patterns with the conversational agent can provide insights into the client's characteristics: \qt{Initially, C7 shared little self-disclosure with the AI, and even now, deep self-disclosures remain limited. This pattern may offer insights into her personality and interpersonal relationships~(T5)}.

\subsubsection{\textbf{Therapists used homework history to trigger discussions in one-on-one sessions}}
We found that AI compiled homework history can help identify meaningful discussion triggers for one-on-one sessions, in three themes:

\textbf{Therapists initiated discussions about the usage of AI brush objects.} 
T1, T2, T3 and T5 noted that they initiated discussions about the usage of AI brush objects selected by clients during in- sessions.
For instance, C9 completed her homework showing two people on opposite sides of a bridge, about to part ways. T5 used \qt{Bridge} as triggers during the in-session: \qt{I hoped she would share something proactively, but she did not. I noticed that the bridge made her feel sad, so I thought the bridge as a chance to discuss}.
Also, T1 and T3 explained that they initiated discussion about the recurring brush objects in artworks, as summarized by the AI: \qt{C1 often included dark clouds in her artwork, and you can see these recurring elements through the AI summary. The recurring brush elements prompted further discussion around these elements~(T3)}. 
T1 added that: \qt{I didn't make immediate judgments, but I was curious about the meaning of elements that frequently appear, as I believed they hold symbolic significance}.

\textbf{Therapists used homework dialogue history data to initiated conversations.} 
T1 and T3 suggested that they used homework dialogue history data to explore topics in depth. 
For example, C17 created an artwork with \textit{Television} and \textit{Media} brushes. 
The homework dialogue history data revealed how the external world overwhelmed him, potentially causing a loss of identity. 
T1 noticed these feelings as triggers for online session discussion: \qt{I have some questions to discuss with C17: what is his true self like? And how has the outside world impacted him?}.
T3 noted that she employed the dialogue data to learn about C5's experiences, enabling more in-depth discussions during in-sessions: \qt{He expressed his desire to go home to honor his mother and shared a lot with the AI. In our second online session, I didn't let him continue to create because his story already contained many aspects that needed to be explored in depth}.

\textbf{Therapists  
initiated discussions about the clients' usage patterns of the system} 
T4 asked clients about their feelings based on the clients' usage patterns of the system that day: \qt{I noticed that a client created four artworks in one day...I would ask if anything noteworthy happened or what inspired their creativity on that day}.
T3 noted that it might capture clients' emotional fluctuations and makes them feel valued: \qt{I used the report as an icebreaker. I remember C13 creating three paintings that day, including a mountain with a forked path and an amusement park, which might have been influenced by certain emotions or events. When I asked about it, she opened up about the career pressures she was facing, feeling truly heard and supported}.

       
\subsubsection{\textbf{Therapists employed homework outcomes to positively influence clients}}
In our study, we discovered that our therapists utilized AI-complied therapy homework outcomes as empowering resources for clients in two key ways:

\textbf{Therapists used homework outcomes to help clients identify their strengths.} 
T2 and T3 mentioned that they used AI-compiled homework outcomes strength-based resources during in-sessions, helping empower clients to recognize and build upon their strength.
For example, T3 employed a homework image as intervention resources to encourage C3 to transform positive feelings: \qt{During the online session, an artwork she created on the spot conveyed an unhappy state. So, we explored resources in her life that could evoke positive emotions. Eventually, we found the homework image of a dog. I hope that by comparing the two images, she might gain a new perspective}. 
C3 noted that: \qt{While discussing family issues, I created a fire to express my anger, unsure how to extinguish it. T3 then shared my previous artwork, asking if it made me feel any better. That’s when I realized there were things that could nourish me with positive energy. In the one-on-one session, the artwork felt like the perfect touch}.
T2 used the clients' artworks to help them discover their strengths: \qt{That day, I told C21 it was our final session. I connected the images from his previous family-themed homework, including childhood moments of playing with his family, and noticed a strong sense of continuity. I highlighted his strengths using specific examples}.
Therapy homework as empowering resource might enable clients to focus on positive aspects and recognize their own strengths, thereby fostering greater self-confidence~\cite{tanner2016homework}.

\textbf{Therapists utilized homework outcomes for transformative reprocessing.} 
T4 mentioned that she repurposed homework artworks as therapeutic tools, guiding clients to reinterpret or modify original creations: \qt{I might download her artwork and show it to the client, asking if she would like to add any elements or rotate different angles to make the artwork better align with her expectations}. 
C11 reflected that: \qt{I draw a bouquet of flowers suspended in mid-air, detached from the soil. T4 showed me the image upside down, asked how I felt, and to give it a new name. I told her the flowers seemed about to fall, and I wanted to anchor them into the soil}.
Also, she tended to revisit and revise images based on her homework artwork: \qt{I would like to help the clients review the last artwork they created, or the one that left the deepest impression on them, and encourage them to create again on the same piece...Recreating the same artwork can help them approach the issue from different perspectives}.
The transformative reprocessing might offer an opportunity to re-examine and transform self-perception, encouraging clients to shift their understanding of the images~\cite{brockway2019art}.

\subsubsection{\textbf{Risks and concerns}}
Although tracking clients' homework history offers many benefits, there are also some concerns and risks associated with it.

\textbf{Tracking homework history adds to therapists' workload.} 
While AI summary save a lot of time, they still spent a considerable amount of time reviewing homework history, potentially adding to their workload. 
For example, T4 complained that: \qt{While the AI summary could save me some time, I used this system to review so many homework sessions that initially required significant effort. For example, the clients might use it multiple times a day, and though a single session might seem to take just 10 minutes, reviewing each session actually takes much longer}.

\textbf{Lacking emotional and behavioral data about clients' creative process.} 
T1 and T4 suggested that the art therapy homework history lacks emotional and behavioral data about the clients' creative process. 
For example, T4 noted that: \qt{In art therapy, we focus on clients' immediate state—not just your words, but also your tone, facial expressions, response time, and overall demeanor. These subtle changes can reveal much about your emotional state, but they are often hard to capture in homework records, limiting a fuller assessment of your needs}. T1 explained that without the emotional and behavioral data, assessments might be incomplete: \qt{If we only use it for pre-assessment, I might feel it's not comprehensive enough. For instance, it misses out on the client's physical gestures, which are crucial in art therapy as we need to assess the entire process}.

\textbf{AI summary sometimes might be misleading.} 
We intentionally designed the AI summary assistant prompts to avoid interpreting or inferring from homework history data, focusing instead on generating natural language summaries and providing relevant insights. But, T2 and T3 found that AI summary might sometimes mislead the therapists. For example, T2 appreciated the AI summary for providing valuable insights, but mentioned that sometimes, \qt{The AI summarized her artwork as depicting a beautiful summer and comparing the can to a `canned universe', but ultimately interpreted it as symbolizing the client's desire to escape reality. This interpretation is misleading and could easily cause confusion}.

\section{Discussion}
Art therapy homework often integrates art-making with verbal expression to explore deep thoughts and uncover new meanings~\cite{hoshino2011narrative,smith2019visual}.
However, the threshold for both accessing and creating meaningful therapy homework is high~\cite{du2024deepthink,Oewel_2024}. 
Further, promoting deep self-reflection during homework can be challenging without the guidance of therapists~\cite{kim2024mindfuldiary}.
Furthermore, another challenge is finding ways to help therapists customize therapy homework and assist them in reviewing and synthesizing homework data during asynchronous collaboration~\cite{Oewel_2024, freeman2007use}.
We thereby set out to explore how to support clients completing art therapy homework in their daily settings~(\textbf{RQ1}) and how to facilitate therapists tailoring and reviewing homework history through asynchronous therapist-client collaboration~(\textbf{RQ2}).

To address this, we have designed and developed \name{}, a human-AI system that integrates co-creative art-making and conversational agents to assist clients in completing art therapy homework. 
Simultaneously, it supports therapists in tailoring the homework agents and reviewing AI-compiled homework history. 
To evaluate \name{}, we conducted a one-month field deployment during which 24 clients used \name{} in their everyday settings, with guidance provided by 5 therapists. 
Addressing RQ1, our quantitative findings revealed that clients used a variety of brushes to engage with their therapy homework over time. Also, 
Our system allowed clients to reflect their emotions and experiences anytime, anywhere in their daily lives, which can offer convenient homework options and lower the threshold of creations.

Prior research has explored how human-AI co-creative art-making within art therapy practices can enhance clients' expressivity and creativity in both synchronous and asynchronous therapy settings~\cite{du2024deepthink}. Further, 
Liu et al. explored combining generative AI with traditional art materials in family expressive arts therapy, where family members guided the process through art and verbal expression, using their creations as inputs for AI-generated images turned into physical materials for storytelling. Extending this line of research, our results found that clients might verbalize emotional feelings and personal experiences during art-making. At the same time, the multi-round human-AI dialogue might can help clients' self-exploration and prompt their interpretation triggered by the unexpected details from the AI-generated images.

In response to \textbf{RQ2}, prior research has highlighted opportunities for AI agents to facilitate asynchronous practitioner-client collaboration in healthcare~\cite{yang2024talk2care}. Conversational agents gathered client health data, which is then summarized and delivered by AI agents to practitioners' interface~\cite{kim2024mindfuldiary}. 
In this line of research, our AI-compiling of homework history, including co-creative artwork, can serve as valuable intervention resources for therapists, helping to encourage clients in transforming and enhancing their positive feelings. 
Meanwhile, inspired by customization of AI agents in HCI~\cite{ha2024clochat,hedderich2024piece}, our findings indicate that therapists incorporate their professional beliefs into conversational agents by modifying conversation principles and example questions to provide structured guidance. Further, therapists infused their personal touch and emotional support into the homework agents through their personal messages.

\subsection{Design Implications}
Below we generalize and discuss a set of design implications to inform future HCI research:

\subsubsection{Implication 1: Combining Multimodal Interaction to Support Therapy Homework via Human-AI Collaboration}
In traditional art therapy homework, integrating art-making with verbalization combines the power of uncovering inner thoughts and finding new meaning through verbalization with the creative and expressive potential of the art-making process~\cite{hoshino2011narrative}.
In our study, our system integrates AI-infused art-making with the collaborative use of conversational agents. The conversational agent captures brush objects from the artwork to facilitate in-depth, multi-turn dialogue with clients. 
Compared to the single, closed monologic forms of therapy homework, the conversational agent allows diverse viewpoints to enter the dialogue and creating a richer, more collaborative conversational texture~\cite{pare2004willow}.
Future designs could combine image segmentation~\cite{kirillov2023segment} and vision-language models~\cite{zhang2024vision} for real-time analysis of artistic elements like colors, shapes, and patterns, and the conversational agent could then adjust its guiding questions to align with clients' emotional state or creative goals.

\textit{Remaining Challenges in the Multimodality of AI-infused Therapy Homework}: Our therapist noted that the AI-infused therapy homework cannot fully replicate the sensory experience associated with using physical art-making materials. 
In art therapy, the kinesthetic and sensory components create a multi-dimensional therapeutic experience that expresses inner sensation and encourages self-soothing~\cite{hinz2019expressive}.
The future therapy homework system could incorporate multimodal design by utilizing human-AI co-creative drawing to generate 3D artwork~\cite{liu2024he}. These AI-generated artworks could be physically printed, offering opportunities to enhance motor-sensory skills while providing calming and therapeutic mental effects.

\subsubsection{Implication 2: Supporting Customizing Well-structured AI Agents as Practitioners' Extension} 
Designing well-structured therapy homework instructions can indeed be challenging for therapists, as they should balance guidance and personal expression to avoid causing confusion.~\cite{Oewel_2024,Harwood2007}. 
Our system allows the therapists to customize the AI agent's conversational principles in one-to-many client therapy homework, enabling it to act as an therapist's extension. T
his provides flexible guidance and questioning, and real-time emotional support during doing therapy homework.
Future work can focus on developing AI agents that build therapist corpora and knowledge bases to continuously empower therapists' practice rather than replace them. 
For example, designing teachable AI agents (similar to apprentice roles~\cite{chhibber2021towards}), learning therapists' language styles and guidance skills to build robust corpora and knowledge bases.

\textit{Risks and Cautions Regarding for Therapist-Customized Therapy Homework Agents:} We found that that as an extension of the therapists, clients may develop overly high expectations of the AI, such as seeking treatment advice from it. 
It may lead to reduced vigilance and overreliance, potentially causing patients to blindly accept incorrect advice~\cite{mendel2024advice}.
Future designs could incorporate a collaborative mechanism between AI agents and therapists. When users actively seek medical advice from the AI agents, it could guide them to connect with a licensed art therapist for further consultation.
We also recommend conducting pre-AI deployment training when deploying the AI-infused art therapy homework system.

\subsubsection{Implication 3: Supporting Longitudinal Multidimensional AI Summary to Simplify Therapists' Homework Tracking.}
Therapists often struggle to easily access and adequately track therapy homework history, resulting in a substantial increase in their workload.
To address this challenge, we leveraged the capabilities of LLMs to help summarize the interactions between clients and AI agents, including artwork descriptions and homework dialogue history.
The AI's summarization can quickly help the therapists gain insight into the client's experiences and highlight recurring brush elements.
Future work could summarize these co-creative and homework dialogue history from more dimensions. 
For example, we could incorporate AI pose recognition (e.g., OpenPose) and facial expression analysis (e.g., Affectiva) to capture detailed posture and expression changes during therapy homework. 
The data, combined with LLMs, could enable cross-analysis of gestures, dialogues, and the artworks.
Further, it is suggested that synthesizing long-term homework history data could provide therapists with valuable insights into clients' personal growth over time~\cite{yang2024talk2care}. 
For instance, we can collaborate with therapists to identify trends and changes in clients' homework history over different time periods, enabling a comparison with their current progress and status.

\textit{Risks and Concerns of AI Summary:} 
Our therapists expressed concerns about the AI summary's reliability, due to occasional inaccuracies or over-speculation. 
Therefore, the therapists hoped to enhance the accuracy and reliability of the AI summary by incorporating more robust theoretical support. 
It is recommended to integrate a knowledge base of art therapy and therapy homework theories into the AI's summarization, enabling therapists to select specific theoretical models for generating summaries.
To enhance the transparency and interpretability of the AI summary, we can incorporate correctness likelihood strategies~\cite{ma2023should} and provide explanations of the theoretical foundations behind the summary, aiding therapists in making informed decisions.

\subsection{Limitations and Future Work}
Our study faces several limitations:
First, we chose to use commercial APIs for conversation agent development, which may raise concerns about user data privacy.
Especially for LLM-driven healthcare systems, which may collect sensitive personal information, there is a risk of medical data leakage~\cite{yang2024talk2care}. 
Future designs could explore deploying localized AI models on edge devices, avoiding centralized server uploads. Additionally, during deployment, users should be well-informed about data collection practices. Regularly publishing privacy policies and data usage reports can enhance transparency, while allowing users to view, modify, and delete their data builds trust.

Another important aspect regards the safe use of generative AI. 
To minimize discomfort or uncanny effects, we incorporated predefined prompts as scaffolding to better control the resulting image, e.g., we included predefined prompts to add characters with Asian facial features, ensuring that clients would create characters that align with their expectations.
However, generative AI occasionally generated images that did not fully meet clients' expectations, potentially causing some tasks frustration.
Future work should prioritize collaboration with art therapists to define the model's application scope and safety boundaries. 
Also, developing a specialized dataset designed for therapeutic purposes—incorporating artworks that reflect emotional expression or commonly used therapeutic symbols—could significantly improve the model's effectiveness in supporting therapy homework.

Finally, increasing the sample size would enhance the validity of the results. Future work should aim to include more participants in extended longitudinal studies to further investigate the generalizability of \name{} in real-world applications.

\section{Conclusion}
In this study, we set out to empirically explore how AI agents support art therapy homework and facilitate therapist-client collaboration. In doing so, We developed and implemented \name{}, a human-AI system that includes a client-facing application for co-creative art-making and conversational agents to assist with therapy homework, as well as a therapist-facing application that allows for customization of homework agents and access to AI-compiled homework history. A one-month field study involving 24 clients and 5 therapists demonstrated how \name{} facilitated clients' homework and reflection through drawing while speaking in their natural environment.
Also, our results explored how therapists infused their practical experiences and personal touch into the agents to customize the homework, and how AI-compiled homework history became practical resources for in-session interactions.
Based on the rich data, we generalize relevant design implications to inform future work in better supporting asynchronous client-practitioner collaboration through human-AI interactions.

\bibliographystyle{ACM-Reference-Format}
\balance
\bibliography{reference.bib}

\newpage
\appendix
\section{Appendix}

\subsection{Conversational agent prompts for generating stable diffusion prompts in art-making phase}

\textbf{Role:} You will be able to capture the essence of the sessions and drawings in the recordings based on the art therapy session recordings I have given you and summarize them into a short sentence that will be used to guide the PROMPT for the Stable Diffusion model.

\vspace{0.5em} 

\textbf{Example input:}

\begin{itemize}[leftmargin=*]
    \item \textbf{USER:} [user-drawn] I drew the ocean. [canvas content] There is nothing on the canvas right now.
    \item \textbf{ASSISTANT:} What kind of ocean is this?
    \item \textbf{USER:} [user-drawn] I drew grass. [canvas content] Now there is an ocean on the canvas.
    \item \textbf{ASSISTANT:} What kind of grass is this?
    \item \textbf{USER:} [user-drawn] I drew the sky. [canvas content] Now there is grass and ocean on the canvas.
    \item \textbf{ASSISTANT:} What kind of sky is this?
    \item \textbf{USER:} [user-drawn] I drew mountains. [canvas content] Now there is sky, grass, and ocean on the canvas.
    \item \textbf{ASSISTANT:} What kind of mountain is this?
    \item \textbf{USER:} [user-drawn] I drew clouds. [canvas content] Now there is sky, mountain, grass, and ocean on the canvas.
    \item \textbf{ASSISTANT:} What kind of cloud is this?
    \item \textbf{USER:} [user dialogue] Colorful clouds, emerald green mountains and grass, choppy ocean
\end{itemize}

\vspace{0.5em} 

\textbf{Task:}

\begin{enumerate}[label=\textbf{Step \arabic*:}]
    \item \textbf{[Step 0]} Read the given transcript of the art therapy session, focusing on the content of \texttt{user: [user drawing]} and \texttt{user: [user dialog]}: Go to \textbf{[Step 1]}.
    \item \textbf{[Step 1]} Based on the input, find the last entry of user's input with \texttt{[canvas content]}, find the keywords of the screen elements that the canvas now contains (in the example input above, it is: sky, grass, sea), separate the keywords of each element with a comma, and add them to the generated result. Examples: [keyword1], [keyword2], [keyword3], \dots, [keyword n].
    \item \textbf{[Step 2]} Find whether there are more specific descriptions of the keywords of the painting elements in \texttt{[Step 1]} in \texttt{[User Dialog]} according to the input. If not, this step ends into \textbf{[Step 3]}; if there are, combine these descriptions and the keywords corresponding to the descriptions into a new descriptive phrase, and replace the previous keywords with the new phrases. Examples: [description of keyword 1] [keyword 1], [keyword 2 description of keyword 2], [description of keyword 3], \dots. Based on the above example input, the output is: rough sea, lush grass, blue sky.
    \item \textbf{[Step 3]} Based on the input, find out if there is a description of the painting style in the \texttt{[User Dialog]} in the dialog record, and if there is, add the style of the picture as a separate phrase after the corresponding phrase generated in \texttt{[Step 2]}, separated by commas. For example: [description of keyword 1] [keyword 1], [description of keyword 2] [keyword 2], \dots, [screen style phrase 1], [screen style phrase 2], [screen style phrase 3], \dots, [Picture Style Phrase n].
\end{enumerate}

\vspace{0.5em} 

\textbf{Output:} 

Only need to output the generated result of \textbf{[Step 3]}.

\vspace{0.5em} 

\textbf{Example output:} 

\emph{Rough sea, lush grass}

\subsection{Conversational agent prompts for discussion phase}

\textbf{Role:} <therapist\_name>, Professional Art Therapist

\textbf{Characteristics:} Flexible, empathetic, honest, respectful, trustworthy, non-judgmental.

\vspace{0.5em} 

\textbf{Task:} Based on the user's dialogic input, start sequentially from step [A], then step [B], to step [C], step [D], step [E] \dots Step [N] will be asked in a dialogical order, and after step [N], you can go to \textbf{Concluding Remarks}. You can select only one question to be asked at a time from the sample output display of step [N]! You have the flexibility to ask up to one round of extended dialog questions at step [N] based on the user's answers. Lead the user to deeper self-exploration and emotional expression, rather than simply asking questions.

\vspace{0.5em} 

\textbf{Operational Guidelines:}

\begin{enumerate}
    \item You must start with the first question and proceed sequentially through the steps in the conversational process (step [A], step [B], step [C], step [D], step [E], \dots, step [N]).
    \item Do not include references like step '[A]', step '[B]' directly in your reply text.
    \item You may include one round of extended dialog questions at any step [N] depending on the user's responses and situation. After that, move on to the next step.
    \item Always ensure empathy and respect are present in your responses, e.g., re-telling or summarizing the user's previous answer to show empathy and attention.
\end{enumerate}

\vspace{0.5em} 

\textbf{Therapist’s Configuration:}

\textbf{Principle 1:}  
\textit{Sample question:} How are you feeling about what you are creating in this moment?

\vspace{0.5em}

\textbf{Principle 2:}  
\textit{Sample question:} Can you share with me what this artwork represents to you personally? 

\vspace{0.5em}

\textbf{Principle 3:}  
\textit{Sample question:} When you think about the emotions connected to this drawing, what comes up for you?

\vspace{0.5em}

\textbf{Principle 4:}  
\textit{Sample question:} How do you connect these feelings to your experiences in your daily life?

\vspace{0.5em} 

\textbf{Concluding Remarks:} Thank participants for their willingness to share and tell users to keep chatting if they have any ideas

\vspace{1em} 

\textbf{Output:} Thank you very much for trusting me and sharing your inner feelings and thoughts with me. I have no more questions, so feel free to end this conversation if you wish. Or, if you wish, we can continue to talk.

\subsection{AI summary prompts}

\textbf{Role:} You are a professional art therapist's internship assistant, responsible for objectively summarizing and organizing records of visitors' creations and conversations during their use of art therapy applications without the therapist's involvement, to help the art therapist better understand the visitor. At the same time, this process is also an opportunity for you to ask questions of the therapist and learn more about the professional skills and knowledge of art therapy.

\textbf{Characteristics:} Passionate and curious about art therapy, strong desire to learn, good at listening to visitors and summarizing humbly and objectively, not diagnosing and interpreting data, good at asking the art therapist questions about the visitor based on your summaries.

\textbf{Task Requirement:} Based on the incoming transcript of the conversation in JSON format, remove useless information and understand the important information from the visitor's conversation, focusing primarily on the visitor's thoughts, feelings, experiences, meanings, and symbols in the content of the conversation. Based on your understanding, ask the professional art therapist 2 specific questions based on the content of the user's conversation in a humble, solicitous way that should focus on the visitor's thoughts, feelings, experiences, meanings, and symbols in the content of the conversation. These questions should help the therapist to better understand the visitor, but you need to make it clear that you are just a novice and everything is subject to the therapist's judgment and understanding, and you need to remain humble.

\textbf{Note:} No output is needed to summarize the combing of this conversation.


\end{document}